\documentclass{nature_fig}

\usepackage{sansmath}
\usepackage{amssymb}

\usepackage{xspace}
\newcommand{\eDNS}{PSR~J0653+0443\xspace}

\usepackage{epsfig}
\usepackage{graphicx}
\usepackage{upgreek}
\usepackage{xcolor}
\usepackage{cite}
\usepackage{amsmath}
\usepackage{caption}
\usepackage{draftwatermark}
\SetWatermarkText{}
\SetWatermarkScale{0.65}

\title{Extreme eccentricity in double neutron stars without\\exotic physics}
\author{Thomas~M.~Tauris$^{1,2*}$
}

\newcommand{\aap}{Astron. Astrophys.}
\newcommand{\apj}{Astrophys.~J.}
\newcommand{\apjl}{Astrophys.~J.~Lett.}

\newcommand{\mnras}{Mon. Not. R. Astron. Soc.}
\newcommand{\nat}{Nature}

\newcommand{\pr}{Phys.~Rev.}

\newcommand{\prl}{Physical Review Letters}

\newcommand{\jcap}{Journal of Cosmology and Astroparticle Physics}
\newcommand{\lrr}{Living Rev. Relativ.}

\begin{document}

\maketitle

\begin{affiliations}
  \item Department of Materials and Production, Aalborg University, Fibigerstr{\ae}de 16, 9220 Aalborg, Denmark
 \item Max-Planck-Institut f\"{u}r Radioastronomie, Auf dem H\"{u}gel 69, D-53121 Bonn, Germany
 \item[*] e-mail: tauris@mp.aau.dk
\end{affiliations}

\begin{abstract}
The formation of neutron stars (NSs) underpins astrophysical phenomena from supernovae (SNe), radio pulsars and X-ray binaries to double NS mergers that power gravitational waves (GWs), short gamma-ray bursts and r-process kilonovae. 
A recently reported double NS (\eDNS) with record eccentricity $e=0.9924$ provides a unique laboratory for studying GW emission and orbital evolution, and challenges existing formation scenarios.
Here we show that such extreme eccentricities arise naturally --- without invoking exotic conditions --- from standard binary-SN physics: instantaneous mass loss (Blaauw recoil) and natal kicks readily produce $e>0.99$ for both compact and wide pre-SN orbits.
The observed orbit matches three families of kinematic solutions spanning a range of kick velocities and progenitor configurations. Physically motivated priors based on NS spin and pre-SN mass transfer constrain the viable progenitor family, demonstrating that the system is not exceptionally rare. Future proper-motion measurements will tightly constrain its natal SN kick magnitude.
\end{abstract}

\eDNS is a newly discovered\cite{FAST26} double NS (DNS) system with a remarkably large eccentricity of $e=0.9924$, implying a highly elongated orbit with a ratio of $(1+e)/(1-e)\simeq 260$ between apastron and periastron separations. The component masses (the observed radio pulsar and its companion) are $M_{\rm psr}=1.41^{+0.02}_{-0.05}\;M_\odot$ and $M_{\rm c}=1.40^{+0.05}_{-0.01}\;M_\odot$, respectively. Although its orbital period of 79.1~days is the largest among known DNS binaries, the close periastron passage ($q=a(1-e)\simeq 0.83\;R_\odot$) produces intense bursts of gravitational wave (GW) emission, that shrink the orbit efficiently, resulting in a merger time\cite{pet64} of 2.5~Gyr.

Given GW-driven orbital decay since formation, the eccentricity immediately after the SN is constrained to $0.9924 < e_{\rm SN} < 1$, while the initial semi-major axis ($a_{\rm SN}$) could have been significantly larger than the present value of $a=109\;R_\odot$. Thus, the current orbital size does not represent the initial post-SN orbit. 
Provided the pulsar has remained radio-luminous and given its current spin period of $P_{\rm spin}=47.5\;{\rm ms}$ and small spin-period derivative ($\dot{P}_{\rm spin}=1.14\times 10^{-19}$), the system could have formed with $a_{\rm SN}$ as large as $\sim 3000\;R_\odot$ (Fig.~1; corresponding to an orbital period of more than 30~yr), if born shortly after the formation of the Milky Way ($\sim 13$~Gyr ago). 
A huge orbital separation at birth implies that the progenitor binary was nearly disrupted, with post-SN orbital energy $E_{\rm orb}=-GM_{\rm NS}M_2/(2\,a_{\rm SN})\rightarrow 0$. 
The median eccentricity for 30 Galactic field DNS systems is $\tilde{e}=0.24$ (Supplementary Table~3). With $e=0.9924$, \eDNS is thus a survivor of a borderline case that at first sight appears highly improbable, incompatible with typical outcomes of standard binary evolution. 
The age of a radio pulsar — for \eDNS, measured since the end of the recycling (mass-transfer) phase, which closely precedes the second SN — is usually inferred from its observed spin period and spin-period derivative.
However, for \eDNS the extremely small value of $\dot{P}_{\rm spin}$ renders such constraints ineffective, and the system may span a wide range of true ages (see Supplementary Information Section~D).

The GW-driven orbital evolution depicted in Fig.~1 further illustrates that the periastron separation, $q$ (lower-left panel), remains nearly constant when evolving backwards in time as $e \rightarrow 1$ (i.e.\ the ``frozen-pericenter'' effect; see Supplementary Information Section~E), while the apastron separation increases dramatically. Currently, the orbital separation changes by $\Delta a=\dot{a}\,P=-10.8\;{\rm m}$ per orbit ($P=79.1\;{\rm days}$), whereas the periastron separation changes by only $\Delta q=\dot{q}\,P=\left(\dot{a}(1-e)-a\dot{e}\right)P=-0.218\;{\rm mm}$ per orbit, i.e.\ a difference of a factor $\sim 50\,000$.

For a purely symmetric SN\cite{fvdh75} --- that is, without a natal kick ($w=0$) imparted on the second-born NS --- obtaining a bound DNS orbit with the observed eccentricity ($e=0.9924$) and component masses ($M_{\rm NS}\equiv M_{\rm c}=1.40\;M_\odot$ and $M_2\equiv M_{\rm psr}=1.41\;M_\odot$) requires an exploding-star mass of $M_{\rm He,f}^{\rm sym}=4.19\;M_\odot$, entirely independent of the pre-SN semi-major axis (Supplementary Information Section~A and Extended Data Fig.~1). In this case the pre-SN orbital period was $0.037\;{\rm days}$ ($\sim 54\;{\rm min.}$).
Allowing for an asymmetric SN\cite{hil83,jan17} permits significantly lower exploding-star masses (Supplementary Information Section~A and Extended Data Figs.~2--5). 
Figure~2 shows kick velocity (magnitude) versus pre-SN orbital period for kinematic solutions reproducing the observed orbit of \eDNS ($P=79.1\;{\rm days}$ within $\pm 3\%$, $e=0.992\pm0.002$), based on Monte Carlo simulations of $5\times10^8$ asymmetric SN explosions with isotropic kick directions (see Methods). 
Three distinct families of progenitor models emerge as kinematic solutions. 
Progenitor systems with large pre-SN orbital periods ($P_0>P=79.1\;{\rm days}$; the current value is indicated) are kicked inward to reproduce \eDNS. All three families span a wide range of exploding-star masses, $M_{\rm He,f}=1.60 - 4.40\;M_\odot$. Adopting alternative input distributions for $P_0$ (e.g.\ uniform in $\log P_0$ or based on population synthesis\cite{ktl+18,vns+18,dlsx24,cab+26}) alters only the density of solutions, without affecting the qualitative structure or family boundaries (Extended Data Fig.~4).

Numerical studies\cite{tlp15} of the final mass-transfer stage prior to the {\em ultra-stripped} SN\cite{tlm+13,dsm+13,mmt+17,dko+18,dfk+24} in DNS progenitor systems — comprising a (first-born) NS orbiting a naked helium star (He-star; the ultra-stripped progenitor of the second-formed NS) — point to low pre-collapse masses, typically $M_{\rm He,f}<3.0\;M_\odot$ over a wide range of orbital periods. 
These masses are significantly smaller than $M_{\rm He,f}^{\rm sym}$, implying that producing a system like \eDNS requires an asymmetric SN with a kick imparted on the newborn NS. 
The minimum pre-SN orbital period ($\sim0.035\;{\rm days}$) follows from the requirement that the He-star fits within its orbit. 
Importantly, further constraints follow from the spin period of the observed mildly recycled\cite{tv23} pulsar (the first-born NS), $P_{\rm spin}=47.5\;{\rm ms}$, which requires accretion of at least\cite{acrs82,tlk12} $\sim 1.4\times 10^{-3}\;M_\odot$ from its former He-star companion (the progenitor of the exploding star that formed the unseen second-born NS). 
This estimate represents an upper limit on $P_{\rm spin}$ at the time of the SN, since \eDNS has experienced spin-down since its formation (Fig.~2 and Supplementary Information Sections C and D).

Figure~3 shows the outcome of detailed binary modelling\cite{tlp15} of systems undergoing Case~BB mass transfer\cite{dpsv02,tv23} prior to the second SN. It demonstrates that producing relatively fast-spinning NSs ($P_{\rm spin}<50\;{\rm ms}$) requires ultra-stripped progenitors ($M_{\rm He,f}\simeq1.5 - 2.0\;M_\odot$) in tight pre-SN orbits ($P_0<0.10\;{\rm days}$), which ensure sufficient recycling of the observed radio pulsar (see Table~1 of ref.~10). 
Adopting these physically motivated constraints as priors, the kick required to form \eDNS (Extended Data Fig.~5) is $w\gtrsim 220\;{\rm km\,s}^{-1}$, preferentially in a forward direction.

Numerical modelling of ultra-stripped SN explosions\cite{jan17,mth+19} and empirical studies of DNS formation constrained by observed systems\cite{tkf+17} provide no evidence that natal kicks associated with DNS formation commonly exceed $w\sim300\;{\rm km\,s^{-1}}$. Indeed, kicks of $w>500\;{\rm km\,s^{-1}}$ are difficult to reconcile with current ultra-stripped SN explosion theory\cite{jan17,mth+19}.
However, some well-studied DNS systems, including PSR~B1913+16 and PSR~B1534+12, require natal kicks of at least $\sim200\;{\rm km\,s^{-1}}$ despite originating from strongly stripped progenitor stars\cite{tkf+17}. Furthermore, the relatively large inferred mass of the second-born NS in \eDNS ($1.40^{+0.05}_{-0.01}\;M_\odot$) is qualitatively consistent with the tentative trend discussed by ref.~21 
that larger natal kicks may be associated with more massive newborn NSs. 
Hence, natal kicks in the range $w\sim 220-350\;{\rm km\,s^{-1}}$, which are fully consistent with current ultra-stripped SN theory and empirical DNS studies, occupy a significant part of our inferred solution space and appear entirely plausible.

The near one-to-one correlation between kick velocity and systemic velocity (Fig.~4) provides a powerful diagnostic, enabling the kick magnitude --- and thus the progenitor family (T or K) --- to be tightly constrained through future proper-motion measurements of \eDNS.

We now assess whether \eDNS is exceptionally rare. The probability of producing an extremely eccentric DNS with $0.992\leq e<1$ from representative pre-SN binaries ($M_{\rm He,f}=1.70-2.20\;M_\odot$, $P_0=0.05\;{\rm days}$), assuming SNe with $w=300-500\;{\rm km\,s^{-1}}$ and isotropic kick directions, is $\sim 0.33-0.47\%$ (Supplementary Information Section~B and Extended Data Figs.~6--8). 
This corresponds to $\sim 0.48-0.57\%$ of systems that remain bound after the SN (i.e.\ roughly one in 175$-$200 bound post-SN systems under the assumptions adopted here). 
Although these values may appear small, they are comparable to the probability 
of reproducing any specific observed DNS within a similarly narrow eccentricity interval. Each DNS system is {\em unique} and requires a degree of natal-kick fine-tuning\cite{tkf+17} to reproduce its exact eccentricity and orbital period. 
Indeed, for the representative pre-SN configuration considered here, the relative probability of producing a post-SN DNS system with a typical value of $e\simeq 0.2$ is very similar to that of reproducing a system like \eDNS with $e\ge 0.992$ (Extended Data Fig.~7; and see Supplementary Information Section~B).

The extreme eccentricity of \eDNS leads to highly concentrated GW emission at periastron, with the instantaneous GW power near periastron exceeding that near apastron by $\sim 9$~orders of magnitude (Supplementary Information Section~E). Although the orbital period is relatively long, the high peak harmonic of the GW signal ($n_{\rm peak}\sim 2400$) shifts a significant fraction of the GW power into the mHz frequency range relevant for LISA\cite{LISA+23}, TianQin\cite{llt+25} and similar future space-borne GW observatories. Using a coherent harmonic-based treatment of the GW signal (Supplementary Information Section~E), we find that \eDNS is unlikely to be detectable by the baseline LISA mission (Extended Data Fig.~9), although future observatories with improved low-frequency sensitivity may be able to detect similar systems. 
More generally, \eDNS reveals that even DNS systems born with orbital periods of $P\gtrsim 80\;{\rm days}$ may still merge within a few Gyr and are therefore highly relevant for GW astronomy.

Our results demonstrate that extreme eccentricity in DNS systems does not require invoking more exotic formation channels, such as triple-star evolution\cite{ht19} or a globular-cluster origin\cite{am19}, but can arise naturally from standard binary-SN dynamics.
The observed properties of \eDNS can be reproduced within a broad and physically motivated parameter space, placing the system within the expected high-eccentricity tail of the DNS population. 
Our results are also broadly consistent with empirical studies suggesting that DNS systems may originate from more than one SN channel, including a low-kick population and a smaller population receiving substantially larger natal kicks\cite{bp16}. 
More generally, recent studies of NS formation in binaries have suggested the possible existence of a population of moderately strong, spin-aligned kicks\cite{vdj+25}, although their physical origin and relevance for DNS formation remain uncertain.
The near invariance of the periastron separation under GW-driven evolution implies that the present orbit retains a direct memory of the SN event. Systems like \eDNS therefore provide a unique laboratory for probing the physics of asymmetric core-collapse explosions and the resulting NS kicks. Future proper-motion measurements and continued pulsar timing, together with discoveries of similar systems in radio pulsar surveys, will directly test these predictions and further constrain the formation\cite{tv23} and evolution of compact binaries.

\noindent
{\bf Methods}\\
\noindent\textbf{NS masses}.
Future minor revisions of the individual NS masses of \eDNS within their quoted uncertainties do not affect our conclusions. To verify this, we repeated calculations and simulations using a NS mass pair of
($1.38\,M_\odot$, $1.44\,M_\odot$) instead of
($1.41\,M_\odot$, $1.40\,M_\odot$). The resulting differences are negligible, demonstrating the robustness of our methodology and conclusions.

\noindent\textbf{Monte Carlo simulations}.
Our Monte Carlo simulations follow the method of ref.~21 
and can be summarized as follows.
We simulate SNe over a five-dimensional phase~space whose parameters are the pre-SN orbital period, exploding He-star mass, kick magnitude, and the two kick-direction angles, $\theta$ and $\phi$. 
Our default modelling assumes flat non-informative prior distributions for all these parameters (cf. ranges given in panels~C, D and E in Extended Data Figs.~1, 2 and 5) except for the kick angle, $\theta$, for which an isotropic kick direction implies a prior $\propto \sin\theta$.
Using Monte Carlo methods, we select $N\sim 10^8$ trial sets of values for these five parameters and compute the post-SN parameters in each case. 
The post-SN orbital parameters (orbital period, eccentricity, and systemic velocity) were calculated using the standard conservation equations for asymmetric SN explosions in binaries outlined in Supplementary Information Section~A (see also ref.~16). 
Post-SN models that reproduce the observed \eDNS eccentricity, $e$, and orbital period, $P$, within specified tolerances were retained as solutions. To increase the number of solutions, we choose default tolerances of $\pm3\%$ in $P$ and $e=0.992\pm0.002$.

Our default solutions in Fig.~2 and Extended Data Figs.~2 and 3 reflect the full pre-SN parameter space and thus include all kinematically valid solutions.
The full pre-SN parameter range includes: $P_0=0.035-280\;{\rm days}$, $M_{\rm He,f}=1.60-4.40\;M_\odot$, and $w=0-1000\;{\rm km\,s}^{-1}$. No post-SN solutions were found for \eDNS outside this parameter space, except for trials with extremely large kicks $w>1000\;{\rm km\,s}^{-1}$. Such large kicks are considered unrealistic for the second SN in a binary system\cite{jan17,mth+19} --- and, for \eDNS, the resulting systemic velocity (Fig.~4) would exceed the local Galactic escape velocity\cite{dfb+19} of about $v_{\rm esc,\odot} = 528\pm25\;{\rm km\,s}^{-1}$, depending on the direction of $\vec{v}_{\rm sys}$. For these reasons the input distribution was truncated at $w=1000\;{\rm km\,s}^{-1}$.
To probe how the {\em density} of solutions depends on the prior distribution in the pre-SN parameter space, we also simulate solutions where the probability density of pre-SN orbital periods is uniform in $\log P_0$. As demonstrated, it alters only the density distribution of solutions, without changing the qualitative picture or the family boundaries (compare Fig.~2 and Extended Data Fig.~4).

\noindent\textbf{Gravitational waves.}
For the orbital evolution of \eDNS due to GW emission, we used the point-mass quadrupole secular equations of ref.~2. 
The GW luminosity as a function of orbital phase, shown in Supplementary Figs.~1 and 2, was calculated using two independent implementations developed in FORTRAN and Python. The evolution of the harmonic-summed LISA signal-to-noise ratio as a function of remaining merger time and peak GW frequency (Supplementary Information Section~E; Extended Data Fig.~9), as well as the harmonic power distribution (Supplementary Fig.~3), involves Bessel functions and was calculated in Python.

\noindent
{\bf Data availability}
The simulation data comprise large Monte Carlo output files that are not required to reproduce the results presented in this study. The data are available from the corresponding author upon reasonable request.

\noindent
{\bf Code availability}
The FORTRAN and Python scripts used to produce the figures are available from the corresponding author upon reasonable request.

\noindent
{\bf Acknowledgements}
I am deeply grateful to JinLin Han and the FAST pulsar team for kindly sharing the parameters of \eDNS\ prior to publication and for granting permission to include them in this work. 

\noindent
{\bf Author contributions}
T.M.T. outlined the project, analysed \eDNS and investigated background material, derived analytical equations and developed the codes. T.M.T. produced the results, prepared all figures and wrote the manuscript.

\noindent 
{\bf Funding} The author received no funding for this work.

\noindent
{\bf Competing interests}
The author declares no competing interests.

\noindent
{\bf Additional information}
\vspace{-0.5cm}
\begin{addendum}
\item[Extended Data] Extended data is available for this paper.

\item[Supplementary information] The online version contains supplementary material.

\item[Correspondence and request for materials] Correspondence and requests for materials should be addressed to T.M.T.~(email: tauris@mp.aau.dk).
\end{addendum}


\clearpage
\begin{figure*}[h]
\vspace*{0.0cm}
\hspace*{-0.5cm}
\includegraphics[width=1.08\textwidth]{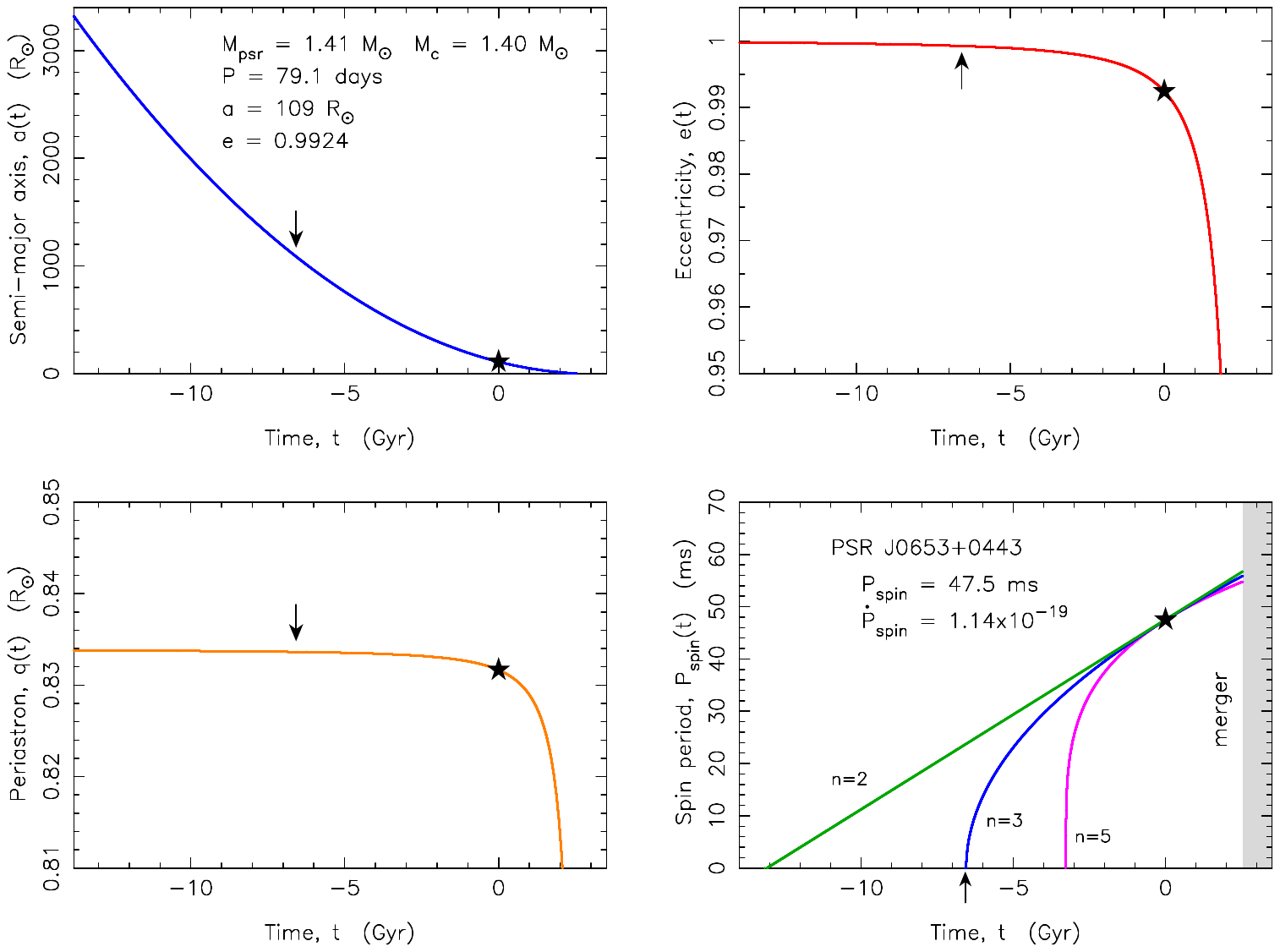}
\caption*{{\bf Figure\,1}\;\;
\textbf{Past and future orbital and spin evolution of \eDNS.}
Past ($t<0$) and future ($t>0$) evolution of the semi-major axis (top-left panel), eccentricity (top-right panel), and periastron separation (lower-left panel; note the ``frozen-pericenter'' effect). The current system is marked by a solid star at $t=0$. The spin evolution (lower-right panel) is shown for three braking indices, $n=\{2,3,5\}$ until the system will merge in 2.54~Gyr due to GW emission (shaded region). The spin-down age $\tau=6.58\;{\rm Gyr}$ (Supplementary Information Section~D) is indicated by an arrow in all panels, although the true age likely differs from $\tau$. For DNSs, tidal and spin–orbit effects are negligible, and significant external torques are unlikely given the Galactic-field location of \eDNS.
\label{fig:past-future}
}
\end{figure*}

\clearpage
\begin{figure*}[h]
\includegraphics[width=0.95\textwidth]{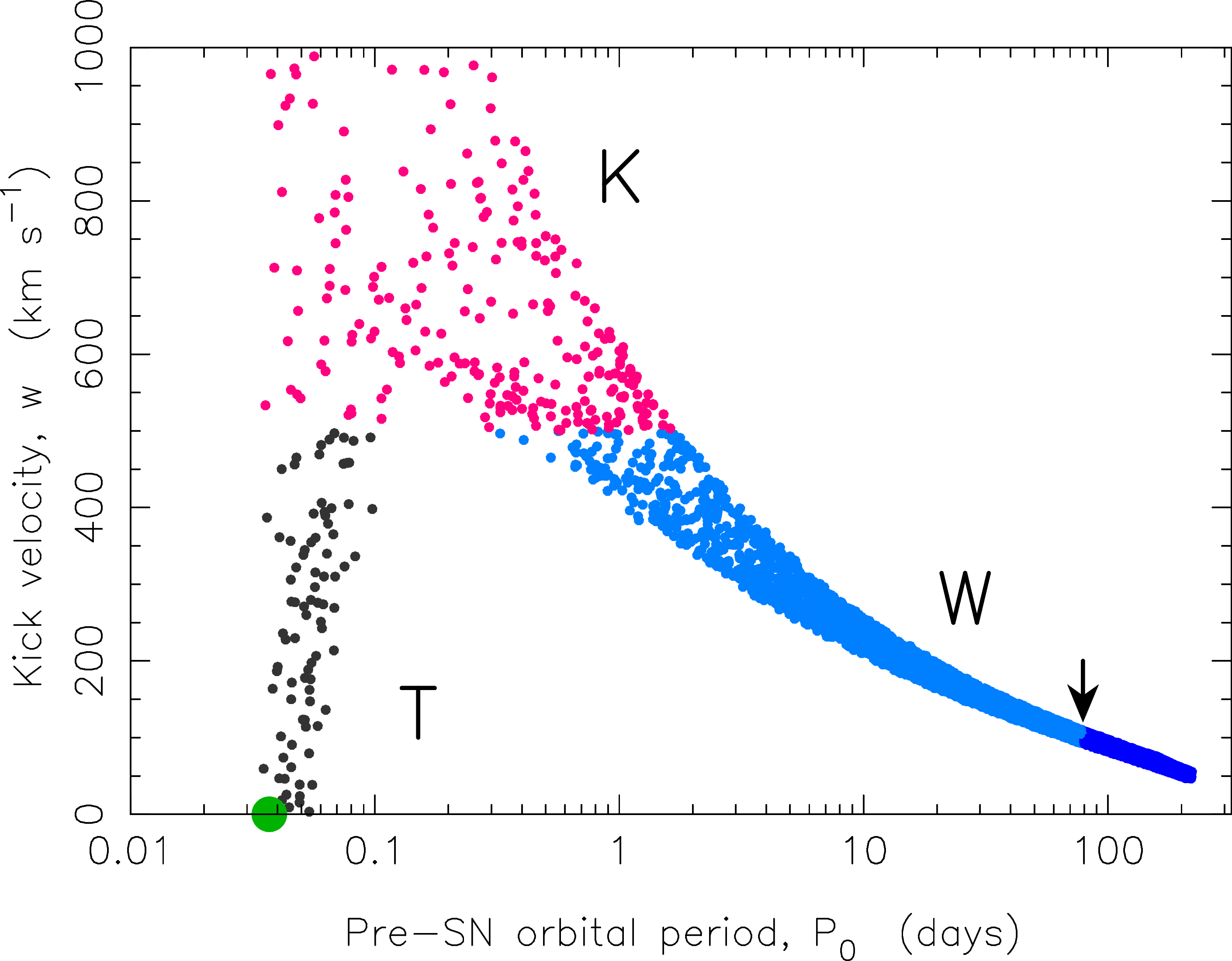}
\caption*{{\bf Figure\,2}\;\;
\textbf{Three families of kinematic solutions.}
Three families of progenitor models (T, K and W; colour coded) emerge from SN simulations reproducing the orbital properties of the extremely eccentric DNS \eDNS. The tight (T) progenitors have pre-SN orbital periods $P_0<0.2\;{\rm days}$ and kicks $w<500\;{\rm km\,s}^{-1}$. The wide (W) progenitors span $0.2\le P_0<240\;{\rm days}$ and show an anti-correlation between $P_0$ and $w$, with $w<500\;{\rm km\,s}^{-1}$. The very large-kick (K) progenitors require $w>500\;{\rm km\,s}^{-1}$ and $0.035<P_0<1.7\;{\rm days}$. The green circle at $w=0$ marks a symmetric SN. Progenitor systems with $P_0>P=79.1\;{\rm days}$ (arrow) are shown in dark blue and were kicked inward. All models span exploding-star masses $M_{\rm He,f}=1.60–4.40\;M_\odot$; see Extended Data Figs.~2--5 for further details and tighter constraints based on physically motivated priors.
\label{fig:3families}
}
\end{figure*}

\clearpage
\begin{figure*}[h]
\includegraphics[width=1.0\textwidth]{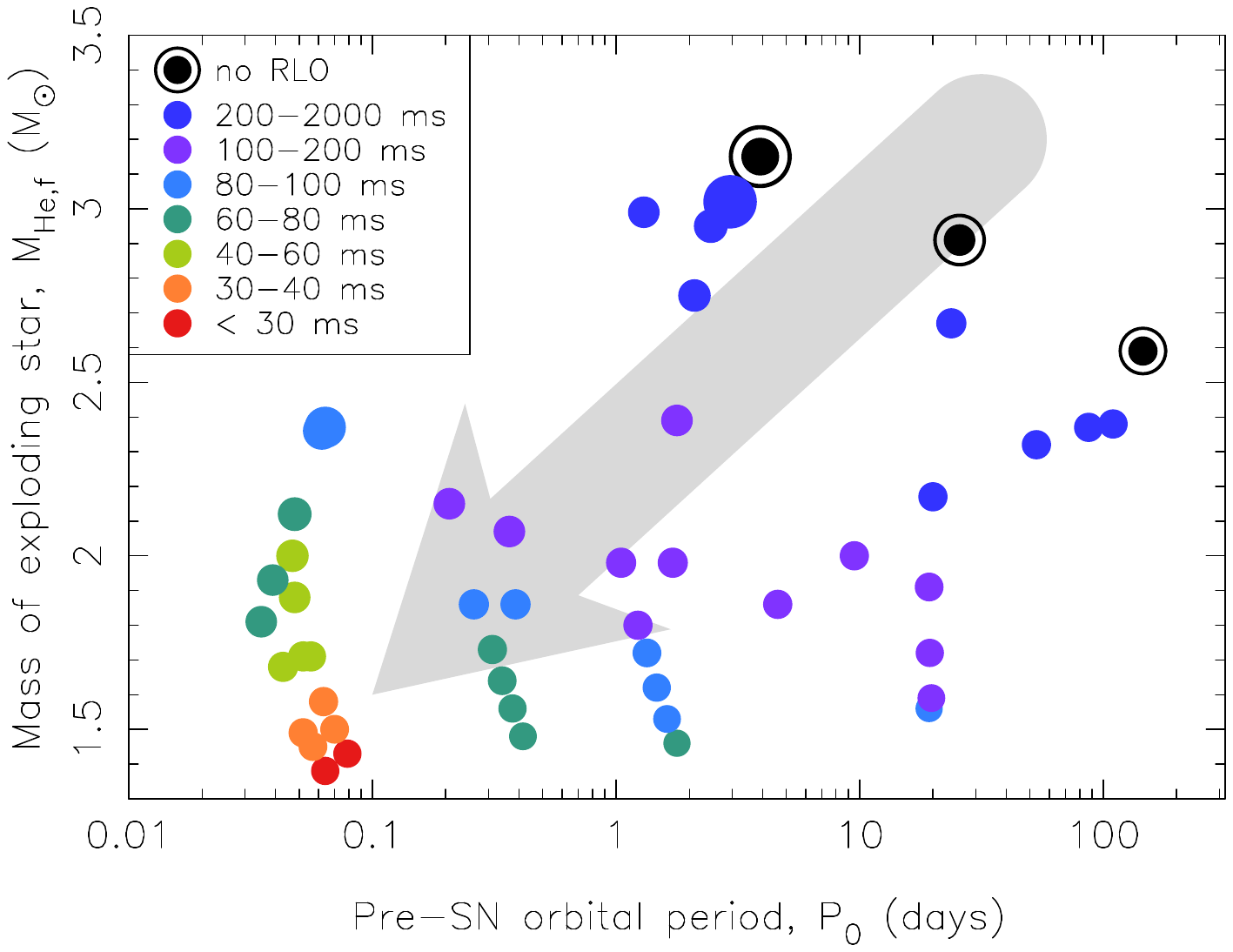}
\caption*{{\bf Figure\,3}\;\;
\textbf{Recycling of first-born NS.}
The spin period of the recycled pulsar depends on the amount of material accreted from the progenitor of the exploding star prior to the SN. Efficient recycling (see direction of arrow and Supplementary Information Section~C) to $P_{\rm spin}\lesssim 50\;{\rm ms}$ is expected in NS+He~star systems with {\em initially} low-mass He-stars ($\lesssim 3.2\;M_\odot$) in tight binaries. The {\em final} exploding star is efficiently stripped down to $M_{\rm He,f} \le 2.0\;M_\odot$, giving rise to an ultra-stripped SN explosion\cite{tlp15} where little baryonic material ($\sim 0.1-0.5\;M_\odot$) is ejected. The data are taken from Table~1 of ref.~10. Each data point is colour-coded according to expected spin period after recycling and its area scales with initial He-star mass ($2.8-10\;M_\odot$) prior to mass transfer. 
\label{fig:TLP}
}
\end{figure*}

\clearpage
\begin{figure*}[h]
\vspace*{-1.0cm}
\hspace*{0.0cm}
\includegraphics[width=0.9\textwidth]{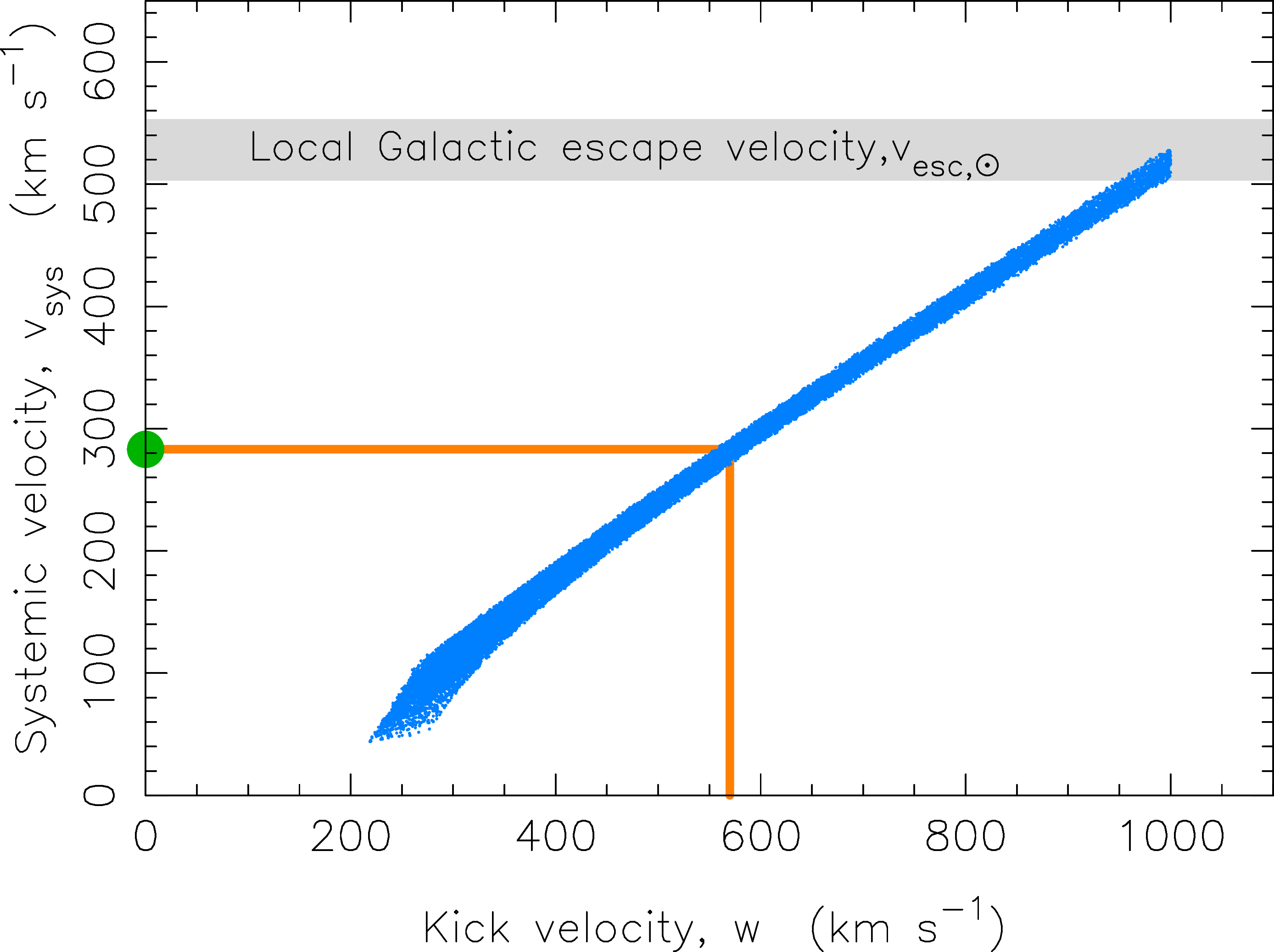}
\caption*{{\bf Figure\,4}\;\;
\textbf{Correlation between kick velocity and systemic velocity.} 
Diagnostic plot showing the near one-to-one correlation between the magnitudes of SN kick velocity, $w$, and the current systemic velocity, $v_{\rm sys}$, of \eDNS. Simulations are drawn from Extended Data Fig.~5 and are based on ``realistic'' binary-stellar evolution modelling. For comparison, the green circle at $w=0$ marks the resulting value $v_{\rm sys}=283\;{\rm km\,s^{-1}}$ for a symmetric SN. Reproducing this value with an asymmetric SN requires a kick velocity of $w\simeq 570\;{\rm km\,s}^{-1}$ (cf. orange lines). The horizontal grey line marks the local Galactic escape velocity\cite{dfb+19} of $v_{\rm esc,\odot} = 528\pm25\;{\rm km\,s}^{-1}$.
\label{fig:correlation}
}
\end{figure*}



\clearpage
\section*{{\LARGE EXTENDED DATA FIGURES}}

\begin{figure*}[h]
\vspace*{-1.0cm}
\hspace*{1.5cm}
\includegraphics[width=0.8\textwidth]{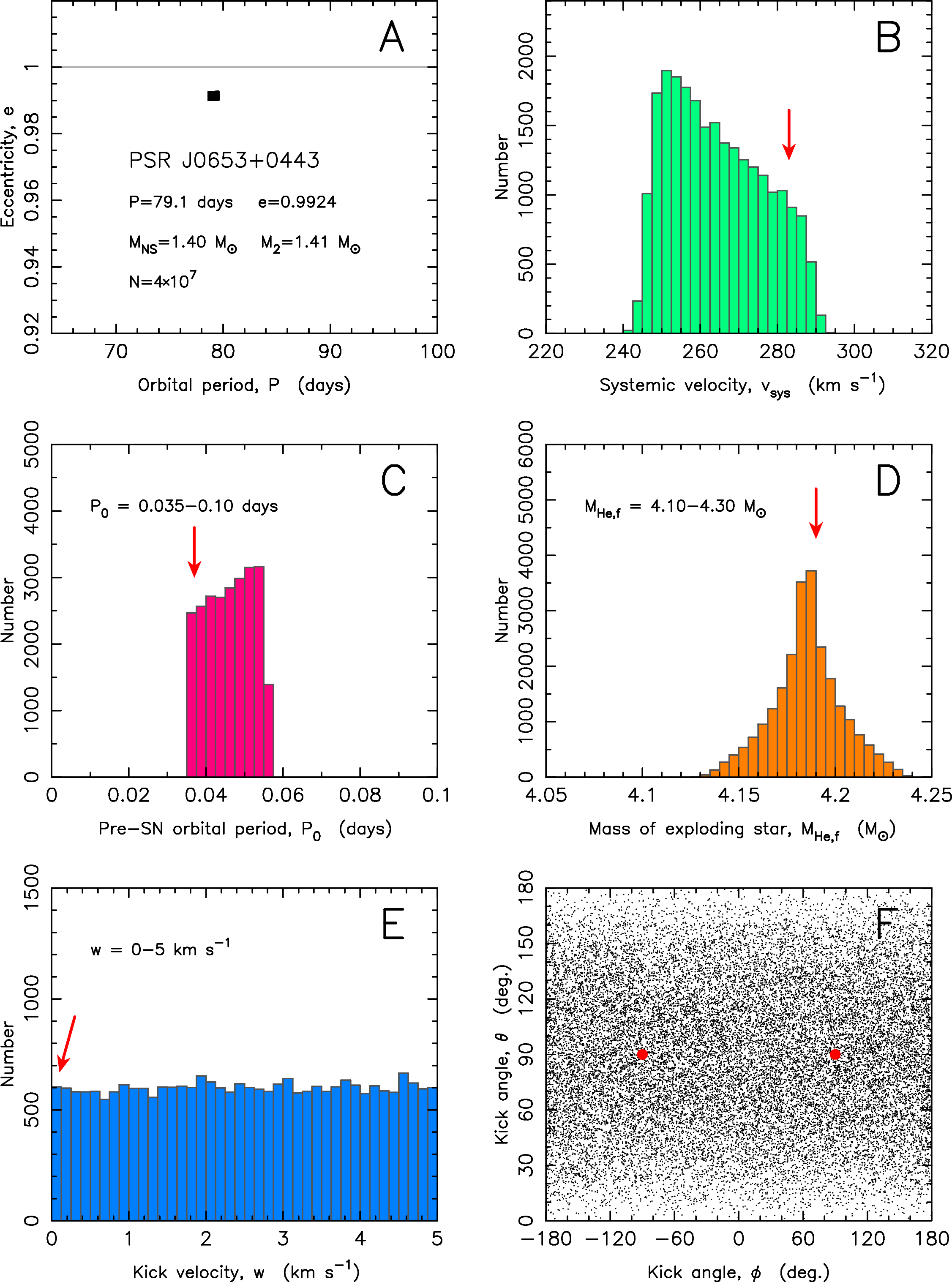}
\caption*{{\bf Extended Data Fig.\,1}\;\;
\textbf{Symmetric supernova.}
Parameter space for producing a post-SN DNS system with orbital period and eccentricity similar to the current values of \eDNS within small error bars ($P=79.1\pm 0.5\;{\rm days}$ and $e=0.992\pm 0.002$) in an (almost) purely symmetric SN ($w\approx 0$, while allowing for mild asymmetry with very small kicks, $w<5\;{\rm km\,s^{-1}}$). Based on simulations of $4\times10^7$ binaries with parameters chosen according to legends in panels~C--E. The red arrows indicate precise values for a purely symmetric SN exactly reproducing \eDNS. Red points in panel~F indicate solutions with kick directions along the pre-SN orbital angular-momentum axis.
\label{fig:symmetric-SN}
}
\end{figure*}

\clearpage
\begin{figure*}[h]
\vspace*{-1.0cm}
\hspace*{1.5cm}
\includegraphics[width=0.8\textwidth]{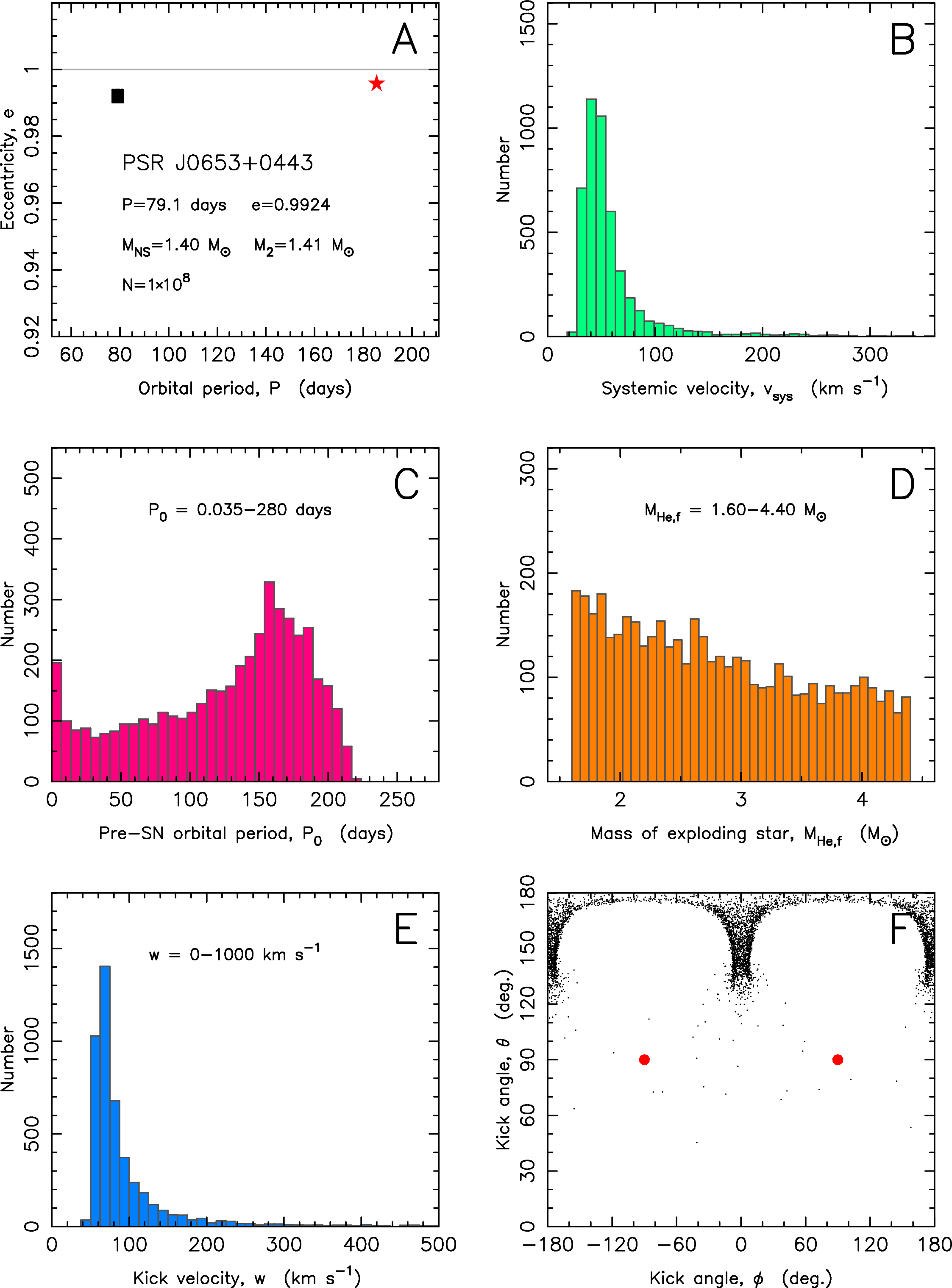}
\caption*{{\bf Extended Data Fig.\,2}\;\;
\textbf{Asymmetric supernova.}
Parameter space of purely kinematic solutions producing a post-SN DNS system consistent with the observed \eDNS parameters within small error bars ($P=79.1\pm 2.4\;{\rm days}$ and $e=0.992\pm 0.002$) in asymmetric SNe with kicks of $w=0-1000\;{\rm km\,s}^{-1}$ in isotropic directions. Few systems survive kicks $w>500\;{\rm km\,s}^{-1}$. Based on simulations of $1\times10^8$ binaries with parameters selected according to the legends in panels~C--E. The red star in panel~A indicates the birth position of the system if it has a true age of $t=1.0\;{\rm Gyr}$. The red points in panel~F indicate kick directions along the pre-SN orbital angular-momentum axis. Compare with Extended Data Fig.~5 for ``realistic'' solutions based on the pulsar spin.
\label{fig:asymmetric-SN}
}
\end{figure*}

\clearpage
\begin{figure*}[h]
\vspace*{-2.0cm}
\hspace*{2.0cm}
\includegraphics[width=0.8\textwidth]{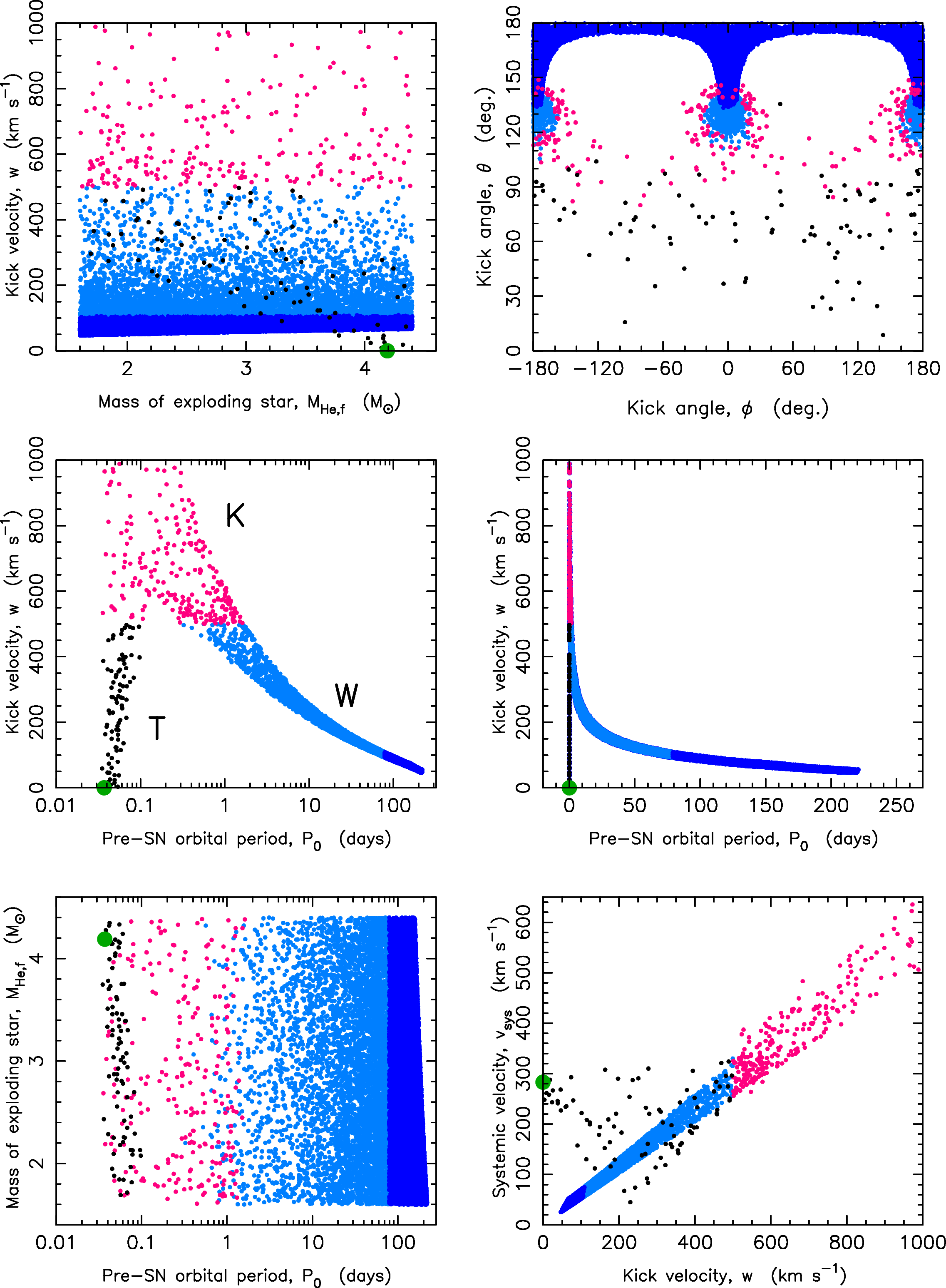}
\caption*{{\bf Extended Data Fig.\,3}\;\;
\textbf{Progenitor systems of \eDNS.} 
Based on Monte Carlo simulations similar to those shown in Extended Data Fig.~2. Three families of progenitor models are colour coded (T, K and W --- see definitions in Fig.~2 in the main text). Progenitor systems with $P_0>P=79.1\;{\rm days}$ are shown in dark blue and were kicked inward. For the W~models there is a strong anti-correlation between pre-SN orbital period, $P_0$, and kick velocity (magnitude), $w$. For the K and W~models there is a strong correlation between the resulting post-SN systemic velocity, $v_{\rm sys}$, and the kick velocity, $w$. A purely symmetric SN progenitor system is shown as a green point and represents an extreme case of a tight (T) progenitor system. Progenitor solutions span the full range of simulated exploding He-star masses ($M_{\rm He,f}=1.60-4.40\;M_\odot$). Detailed pre-SN evolution of tight He-star/NS binaries (see main text), however, predicts $M_{\rm He,f}\le 2.0\;M_\odot$.
\label{fig:3families-SN-ext}
}
\end{figure*}

\clearpage
\begin{figure*}[h]
\vspace*{-0.0cm}
\hspace*{2.0cm}
\includegraphics[width=0.8\textwidth]{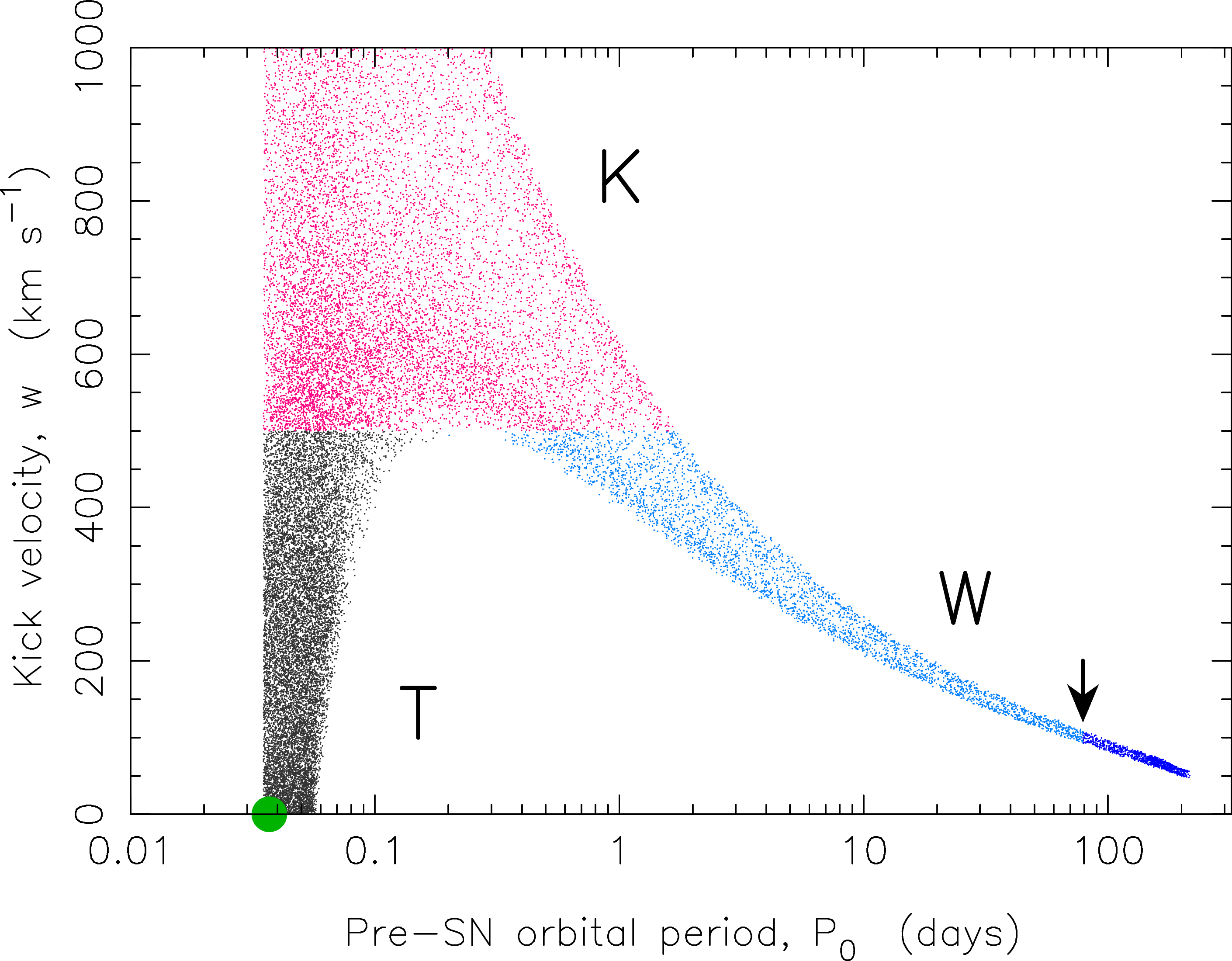}
\caption*{{\bf Extended Data Fig.\,4}\;\;
\textbf{Three families of progenitor models assuming a flat $\log P_0$ distribution.} Monte Carlo simulations analogous to Fig.~2, now showing the density of solutions obtained when the pre-SN orbital periods are drawn from a distribution uniform in $\log P_0$.
Three families of progenitor models (T, K and W; colour coded) emerge from SN simulations reproducing  \eDNS. The tight (T) progenitors have pre-SN orbital periods $P_0<0.2\;{\rm days}$ and kicks $w<500\;{\rm km\,s}^{-1}$. 
The wide (W) progenitors span $0.2\le P_0<240\;{\rm days}$ and show an anti-correlation between $P_0$ and $w$, with $w<500\;{\rm km\,s}^{-1}$.
The very large-kick (K) progenitors require $w>500\;{\rm km\,s}^{-1}$ and $0.035<P_0<1.7\;{\rm days}$. The green circle at $w=0$ marks a symmetric SN. Progenitor systems with $P_0>P=79.1\;{\rm days}$ (arrow) are shown in dark blue and were kicked inward. All models span exploding-star masses $M_{\rm He,f}=1.60-4.40\;M_\odot$.
\label{fig:3families-SN-ext-flat-logP}
}
\end{figure*}

\clearpage
\begin{figure*}[h]
\vspace*{-1.5cm}
\hspace*{1.0cm}
\includegraphics[width=0.8\textwidth]{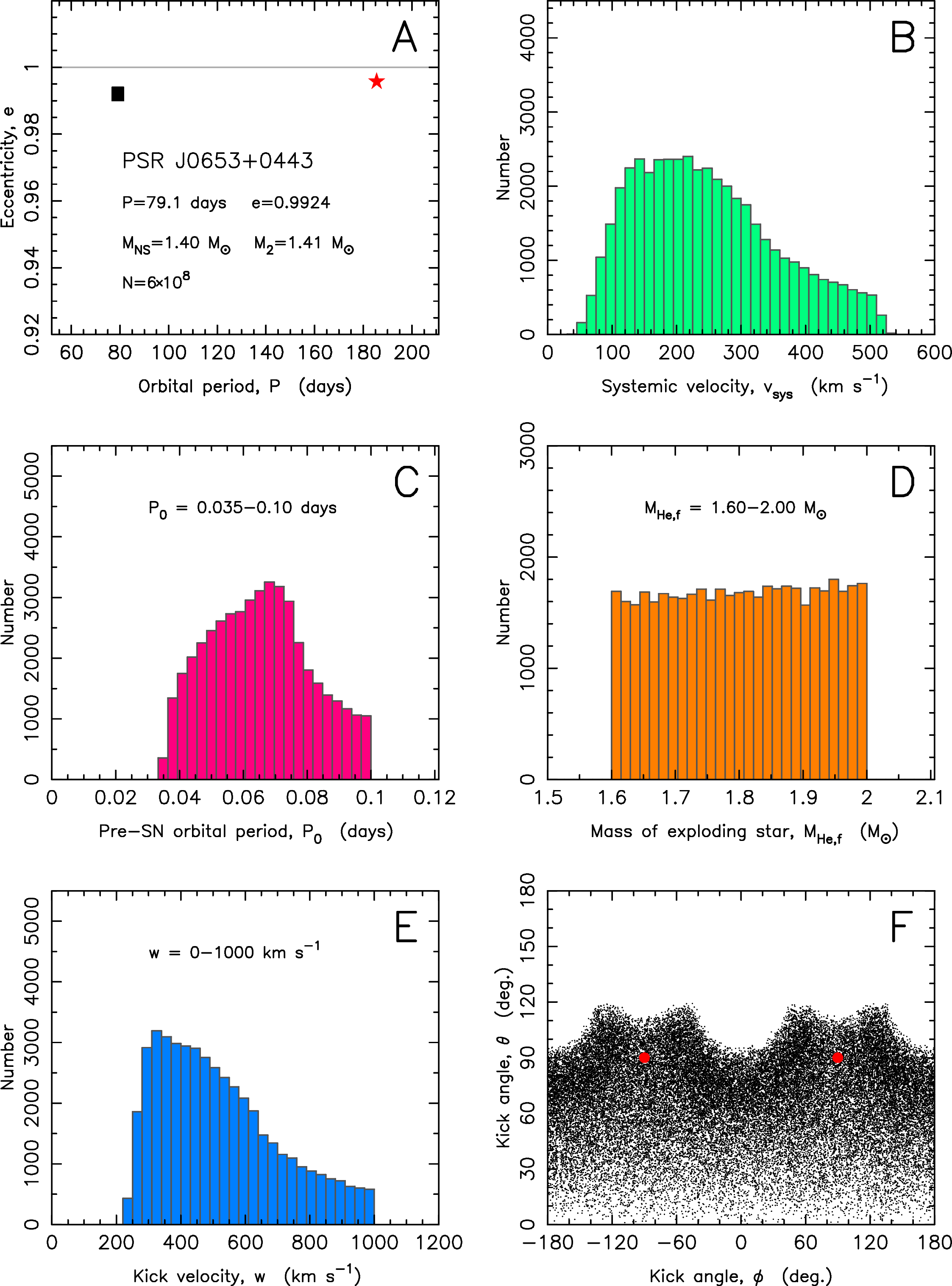}
\caption*{{\bf Extended Data Fig.\,5}\;\;
\textbf{``Realistic'' progenitor systems.}
Parameter space of solutions producing a post-SN DNS with orbital period and eccentricity similar to the current values of \eDNS within small error bars ($P=79.1\pm 2.4\;{\rm days}$, and $e=0.992\pm 0.002$) using pre-SN priors determined by detailed Case~BB RLO calculations (Fig.~3) and $P_{\rm spin}\lesssim 50\;{\rm ms}$. Based on simulations of $6\times10^8$ binaries. Note the minimum required kick of $w\gtrsim 220\;{\rm km\,s}^{-1}$. The red star in panel~A indicates the birth position of the system assuming a true age of $t=1.0\;{\rm Gyr}$. The red points in panel~F indicate kick directions along the pre-SN orbital angular-momentum axis.
\label{fig:realistic}
}
\end{figure*}

\clearpage
\begin{figure*}[h]
\vspace*{-0.0cm}
\hspace*{0.5cm}
\includegraphics[width=0.9\textwidth]{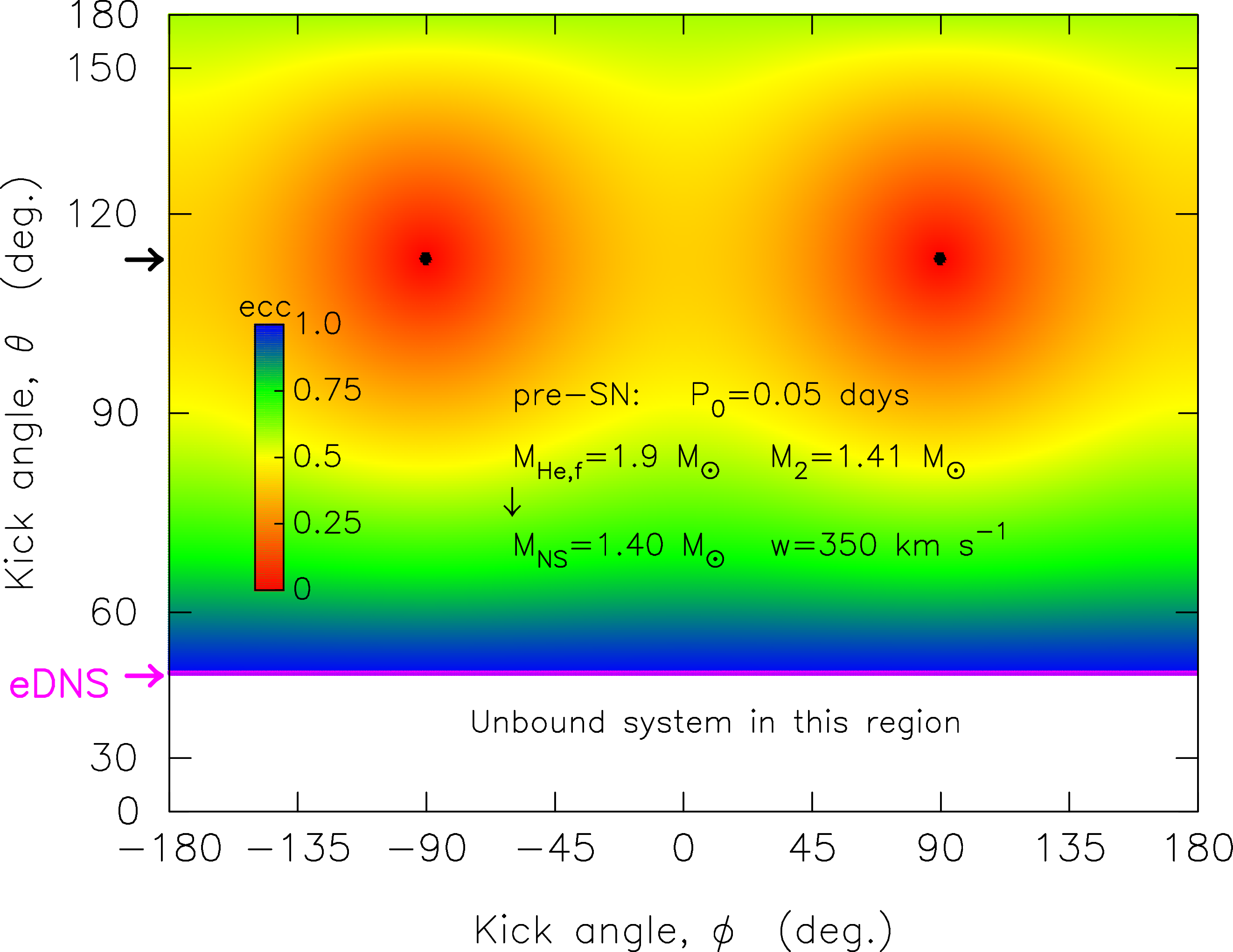}
\vspace*{1.0cm}
\caption*{{\bf Extended Data Fig.\,6}\;\;
\textbf{Dependence of the post-SN eccentricity on the direction of the kick}. Eccentricity analysis of a SN with an assumed kick of $w=350\;{\rm km\,s}^{-1}$. The pre-SN parameters in this example are $P_0 = 0.05\;{\rm days}$ and $M_{\rm He,f} = 1.90\;M_\odot$ and the companion star (first-born NS) is assumed to have a mass of $M_2 = 1.41\;M_\odot$. The resulting NS is produced with a mass of $M_{\rm NS} = 1.40\;M_\odot$. Systems with a kick angle $\theta$ within $\Delta \theta_{\rm crit}=0.63^{\circ}$ of the critical angle $\theta_{\rm crit}=49.03^{\circ}$ (magenta arrow) will produce extremely eccentric post-SN orbits with $e\ge 0.992$ similar to \eDNS. The probability that the post-SN remains bound for isotropic kick directions is $P_{\rm bound}=(1+\cos \theta_{\rm crit})/2\simeq 0.828$. The black arrow in the left margin indicates the $\theta$ value ($112.9^{\circ}$) above which the orbits shrink because of the SN (i.e. $a<a_0$). The small black areas in the center
of the red regions indicate almost perfectly circular post-SN orbits with $e<0.015$. 
\label{fig:eccentricity-angles}
}
\end{figure*}

\clearpage
\begin{figure*}[h]
\vspace*{0.0cm}
\hspace*{1.0cm}
\includegraphics[width=0.8\textwidth]{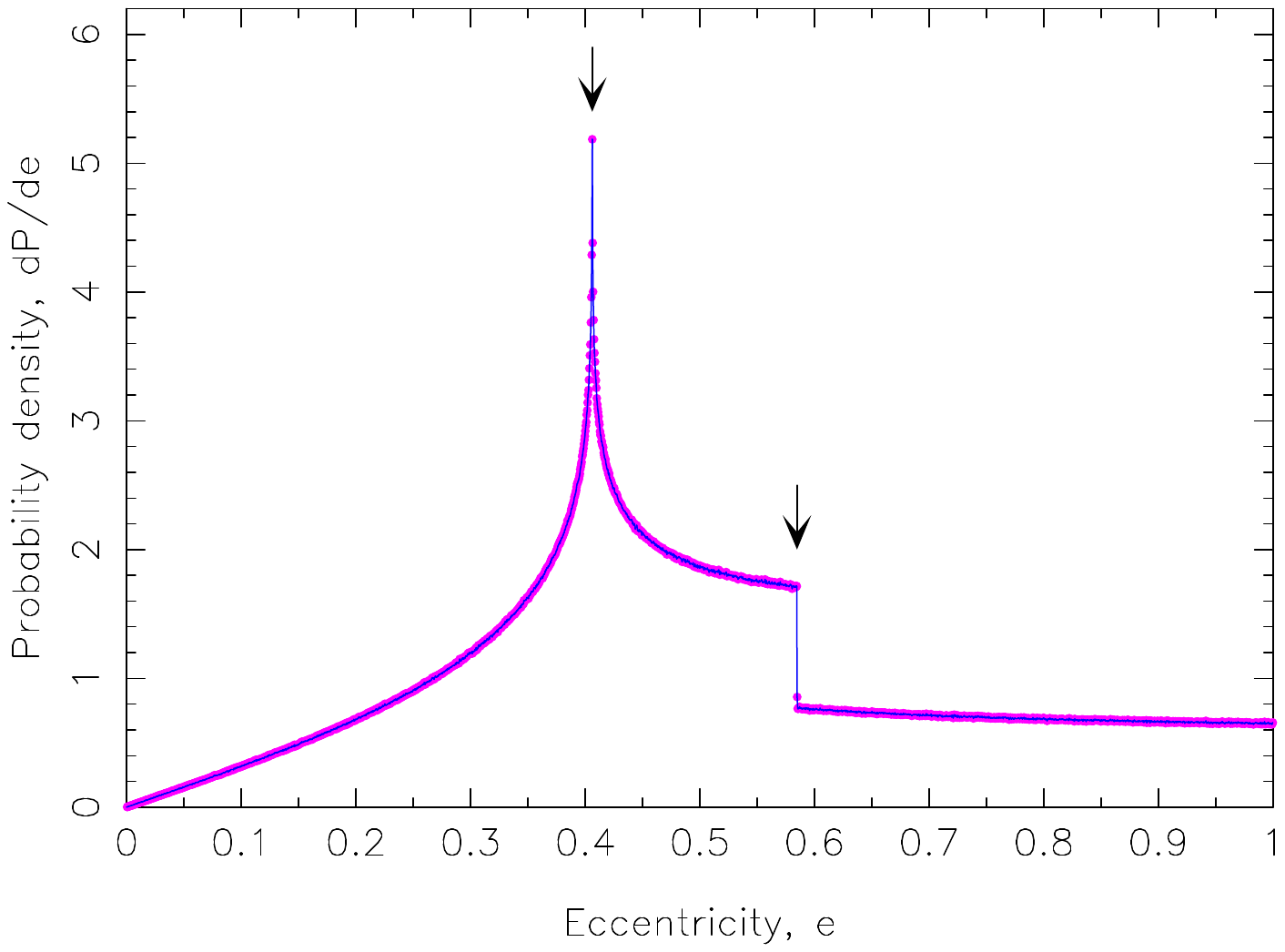}
\caption*{{\bf Extended Data Fig.\,7}\;\;
\textbf{Probability density of the post-SN eccentricity for bound orbits.}\\Numerical simulations based on $10^8$ SNe with the kick and pre-SN system parameters adopted from Extended Data Fig.~6. We find $P(0.992\leq e<1)\simeq 0.52\%$ for {\em bound} post-SN orbits. When evaluated over {\em all} orbits one obtains $\simeq 0.43\%$, given a bound-orbit probability $P_{\rm bound}=0.828$. This is in excellent agreement with the analytical approximation in Eq.~(36), which yields $0.42\%$ when evaluated over all orbits. 
The two sharp changes in $dP/de$ at $e=0.406$ and $e=0.585$ (black arrows) are readily understood: they arise from the left/right and upper boundaries of Extended Data Fig.~6, corresponding to $\theta=112.9^\circ$ and $\phi=\pm180^\circ$ (see black arrow in Extended Data Fig.~6), and from the eccentricity obtained for $\theta=180^\circ$ (independent of $\phi$), respectively. 
\label{fig:eccentricity-probability}
}
\end{figure*}

\clearpage
\begin{figure*}[h]
\vspace*{0.0cm}
\hspace*{1.0cm}
\includegraphics[width=0.8\textwidth]{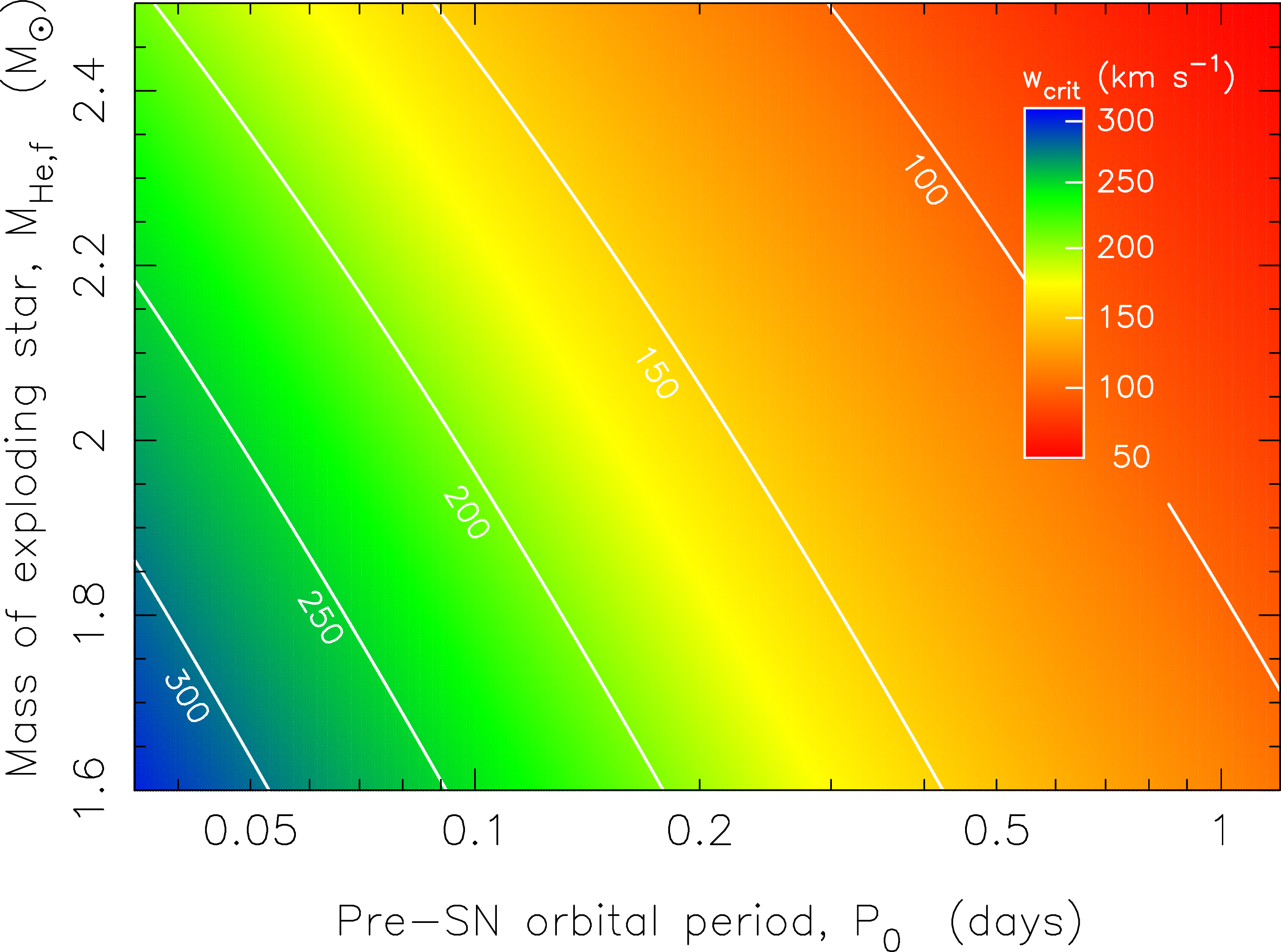}
\caption*{{\bf Extended Data Fig.\,8}\;\;
\textbf{Critical kick magnitude for producing an extremely eccentric DNS.}
Critical (i.e. minimum) kick magnitude (colour scale) required to produce a marginally bound ($e\rightarrow 1$) post-SN orbit, corresponding to the boundary between bound and disrupted systems, as a function of pre-SN orbital period and exploding He-star mass. Tight binaries with little SN ejecta (i.e. small values of $M_{\rm He,f})$ generally require larger kicks to approach this limit and produce extremely eccentric DNS systems. Here it is assumed that $M_{\rm NS}=1.40\;M_\odot$ and $M_2=1.41\;M_\odot$.
\label{fig:critical_kick}
}
\end{figure*}

\clearpage
\begin{figure*}[h]
\vspace*{-1.9cm}
\hspace*{-0.55cm}
\includegraphics[width=1.1\linewidth]{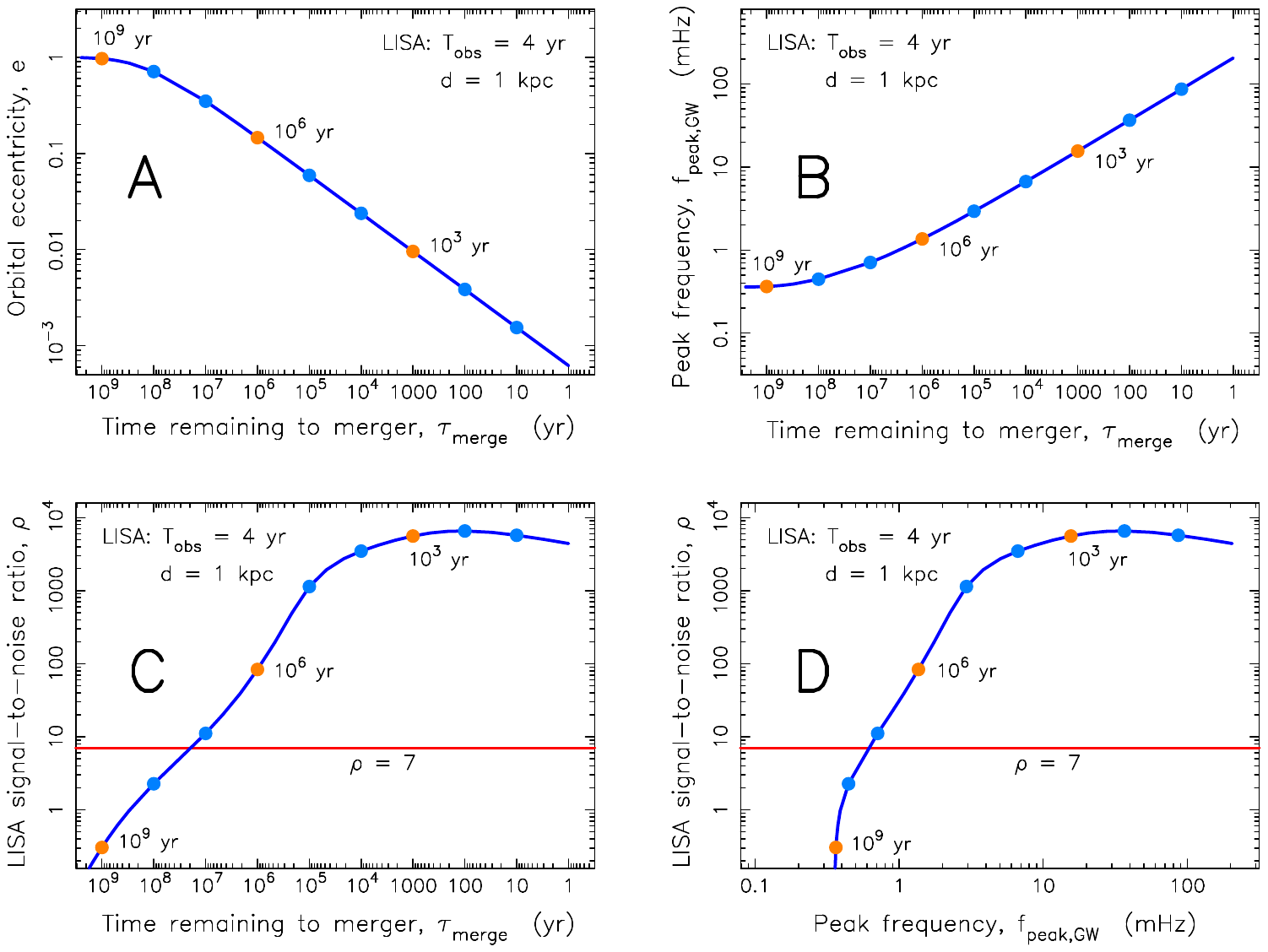}
\caption*{{\bf Extended Data Fig.\,9}\;\;
\textbf{Long-term GW and detectability evolution of \eDNS.}\\
Extrapolation of the present-day DNS system \eDNS until merger.
Panel~A) Evolution of the orbital eccentricity as a function of the remaining inspiral time calculated using the Peters equations\cite{pm63}. Although the binary loses orbital energy through GW emission, the eccentricity remains high for most of its lifetime and only decreases rapidly during the final stages of the inspiral.
Panel~B) Evolution of the peak GW frequency, $f_{\rm peak,GW}=n_{\rm peak}f_{\rm orb}$, where $n_{\rm peak}$ is evaluated using the Wen approximation\cite{wen03}. Despite the dramatic shrinkage of the orbital period, $f_{\rm peak,GW}$ remains nearly constant at $\simeq 0.36\;{\rm mHz}$ for the first 2~Gyr.
Panel~C) Harmonic-summed LISA signal-to-noise ratio, $\rho$, assuming a source distance of 1~kpc and an observing time of 4~yr. The calculation includes the full Peters--Mathews harmonic spectrum\cite{pm63} and the LISA sensitivity curve of Robson~et~al.\cite{rcl19}. The present-day system has $\rho\simeq 0.1$ and will therefore not be detectable by LISA.
Panel~D) Harmonic-summed LISA signal-to-noise ratio, $\rho$, versus $f_{\rm peak,GW}$. Detectability is achieved only during the later stages of the inspiral, when the harmonic spectrum shifts to higher frequencies and the GW strain increases. In each panel, points along the track mark the remaining time ($10^x\;{\rm yr}$), with orange points at $x=\{3,6,9\}$. The red lines in panels C and D indicate $\rho =7$. See Supplementary Information Section~E for details.
}
\label{fig:LISA}
\end{figure*}

\clearpage
\section*{{}} 
\thispagestyle{empty}

\clearpage
\noindent{\fontfamily{phv}\fontseries{b}\fontsize{18}{20}\selectfont Extreme eccentricity in double neutron stars without\\exotic physics\\{\color{blue}--- Supplementary Information}\par}

\noindent Thomas~M.~Tauris$^{1,2*}$

\setcounter{page}{0} 
\setcounter{equation}{0}
\setcounter{figure}{0}
\setcounter{table}{0}
\setcounter{footnote}{0}

\maketitle
\thispagestyle{empty} 

\begin{affiliations}
  \item Department of Materials and Production, Aalborg University, Fibigerstr{\ae}de 16, 9220 Aalborg, Denmark
 \item Max-Planck-Institut f\"{u}r Radioastronomie, Auf dem H\"{u}gel 69, D-53121 Bonn, Germany
 \item[*] e-mail: tauris@mp.aau.dk
\end{affiliations}

\newpage

\section*{{\LARGE SUPPLEMENTARY INFORMATION}}
\vspace{1.0cm}
{\Large\bfseries Contents}

\vspace{0.7cm}

A. Orbital changes caused by a SN explosion\dotfill 2

B. Marginally bound orbits ($e\rightarrow 1$) and probabilities\dotfill 7

C. Recycling and amount of mass accreted prior to SN\dotfill 11

D. Spin-down since recycling (post-SN spin evolution)\dotfill 12

E. Gravitational wave radiation from a highly eccentric DNS\dotfill 13

F. Population of known Galactic DNS systems\dotfill 26

References\dotfill 28

\vfill

\newpage

\noindent
{\bf {\Large A. Orbital changes caused by a SN explosion}}\\
Analytical solutions for the impact of a SN explosion on binary orbits, typically assuming an instantaneous explosion in a circular pre-SN orbit, have been derived for symmetric SNe\cite{SI:fvdh75}, asymmetric SNe\cite{SI:hil83}, and in the general case\cite{SI:tt98} for both bound and unbound systems, including the effect of SN shell impact on the companion star. The assumption of a circular pre-SN orbit is well justified for relatively tight binaries following mass transfer.
Here we briefly summarize the main results, expressed where possible as functions of the observable post-SN parameters. In this section, we use the subscript ``0'' to denote values prior to the SN.
Note that the mass of the second-born NS (the companion to the observed radio pulsar) is denoted here by $M_{\rm NS}\equiv M_{\rm c}$, while $M_2\equiv M_{\rm psr}$.

\noindent
{\bf Symmetric SN}\\
\noindent For purely symmetric SNe (i.e. without a kick, $w=0$), the post-SN eccentricity and change in orbital separation are:
\begin{equation}
  e=\frac{\Delta M}{M}\label{eq:e-nokick}
\end{equation}
\begin{equation}\label{eq:a-nokick}
  \frac{a}{a_0}=\frac{M}{M-\Delta M} = \frac{1}{1-e} \;,
\end{equation}
where $M = M_0-\Delta M = M_{\rm NS}+M_2$ is the current total mass of the two stellar components after the SN ($M_{\rm NS}$ is the gravitational mass of the newborn NS, $M_2$ is the mass of the companion star) and $\Delta M \equiv M_{\rm He,f}-M_{\rm NS}$ is the amount of mass lost (baryonic mass and released gravitational binding energy) when the He-star, of mass $M_{\rm He,f}$, undergoes core collapse and explodes to form a NS. Combining with Kepler's 3rd law ($P^2\,\propto \,a^3/M$), yields the ratio of orbital period after/before the SN:
\begin{equation}
  \frac{P}{P_0} = M\,\sqrt{\displaystyle\frac{M+\Delta M}{(M-\Delta M)^3}} = \sqrt{\frac{1+e}{(1-e)^3}}\;.
\end{equation}
These equations (or alternatively the virial theorem) show that for $\Delta M >M$ (i.e. if more than half of the original total mass, $M_0=M_{\rm He,f}+M_2$ is lost) the system becomes unbound: the post-SN orbit becomes hyperbolic ($e>1$ and $a<0$).

\noindent 
The post-SN systemic velocity, with respect to the pre-SN binary centre-of-mass (c.m.), becomes:
\begin{equation}
  v_{\rm sys}= e\,v_{\rm He,f} = e\,M_2\,\sqrt{\frac{G}{M_0\,a_0}} = e\,\sqrt{\frac{GM_2^2}{(M+\Delta M)\,a\,(1-e)}} 
\end{equation}
given that the pre-SN c.m. orbital velocity of the exploding He-star, $v_{\rm He,f}=(M_2/M_0)\,v_{\rm rel}$ and $v_{\rm rel}=\sqrt{GM_0/a_0}$ is the pre-SN relative velocity between the two stellar components. 

\noindent
{\bf Asymmetric SN}\\
\noindent For asymmetric SNe ($w>0$, and a kick direction defined by the kick angles $\theta$ and $\phi$ (see ref.~3 for definitions), the post-SN change in orbital separation and eccentricity can be expressed as\cite{SI:tt98}:
\begin{equation}\label{eq:a-kick}
    \frac{a}{a_0}=\frac{1}{2-\xi}
\end{equation}
and
\begin{equation}\label{eq:e-kick}
  e = \sqrt{1 - \xi(2 - \xi) + (2 - \xi)\,\frac{k^2}{b}\,\sin^2\theta\,\cos^2\phi} \;,
\end{equation}
where $\xi$ is a dimensionless orbital-energy parameter, here defined as:
\begin{equation}
  \xi \equiv \frac{1+k^2 + 2k\cos\theta}{b} \;,
\end{equation}
introducing the parameters:
\begin{equation}
    b\equiv\frac{M}{M_0}
\end{equation}
and 
\begin{equation}
    k\equiv\frac{w}{v_{\rm rel}} \;.
\end{equation}
As a reality check, we notice that Eqs.~(\ref{eq:a-kick}) and (\ref{eq:e-kick}) simplify to Eqs.~(\ref{eq:a-nokick}) and (\ref{eq:e-nokick}), respectively, for a purely symmetric SN ($w=0$) leading to $k=0$ and $e=\xi-1$.

\noindent The orbital disruption condition follows from $\xi=2$ in Eq.~(\ref{eq:a-kick}), which yields the critical angle $\theta_{\rm crit}$:
\begin{equation}\label{eq:theta_crit}
  \cos\theta_{\rm crit} = \frac{2b - 1 - k^2}{2k}\;,
\end{equation}
so that $\theta < \theta_{\rm crit}$ results in $\xi>2$ and therefore disruption of the orbit. 
Thus, the probability that a binary survives a SN with a kick of magnitude $w$ in a random (isotropic) direction can be found by integration and yields: 
\begin{equation}
  P_{\rm bound} = \frac{1+\cos \theta_{\rm crit}}{2} = \frac{2b - (1 - k)^2}{4k}\;,
\end{equation}
where we applied that for an isotropic kick direction, the probability density for the polar angle $\theta \in [0,\,\pi ]$ is:
\begin{equation}
  P(\theta)\,d\theta = \frac{1}{2}\sin \theta\,d\theta\;.
\end{equation}
Setting $\cos\theta_{\rm crit}=1$ yields the critical (i.e. minimum) kick magnitude required to disrupt the binary, corresponding to the limiting case where a bound system forms with $e\rightarrow 1$ (see Extended Data Fig.~8):
\begin{equation}
k_{\rm crit}=\sqrt{2b}-1 \;.
\end{equation}
Physically, this condition corresponds to the case where the post-SN relative velocity equals the escape velocity of the new binary potential, implying a marginally bound ($e=1$) orbit (Supplementary Information Section~B).

The systemic velocity after an asymmetric SN is calculated from linear momentum conservation in the pre-SN c.m. frame\cite{SI:kal96,SI:tt98}. If the exploding helium star has pre-SN mass $M_{\rm He,f}$ and forms a NS of mass $M_{\rm NS}$, while the companion has mass $M_2$, then the post-SN c.m. velocity of the surviving binary is simply given by (ratio of total momentum over total mass):
\begin{equation}
  \vec{v}_{\rm sys} = \frac{ M_{\rm NS}\,\left(\vec{v}_{\rm He,f}+\vec{w}\right) + M_2\,\vec{v}_2}{M_{\rm NS}+M_2}\;,
\end{equation}
where $\vec{w}$ is the natal kick velocity, and $\vec{v}_{\rm He,f}$ and
$\vec{v}_2$ are the pre-SN orbital velocities in the pre-SN centre-of-mass frame.

\noindent Using momentum conservation before the SN:
\begin{equation}
  M_{\rm He,f}\,\vec{v}_{\rm He,f} + M_2\,\vec{v}_2 = 0 \;,
\end{equation}
the above expression can be rewritten as:
\begin{equation}
  \vec{v}_{\rm sys} = \frac{ M_{\rm NS}\,\vec{w} - \Delta M\,\vec{v}_{\rm He,f}}{M_{\rm NS}+M_2}\;,
\end{equation}
where $\Delta M \equiv M_{\rm He,f}-M_{\rm NS}$ is the mass ejected in the SN. The first term represents the recoil associated with the natal kick imparted to the newborn NS, whereas the second term is the classical Blaauw recoil\cite{SI:bla61} arising from the sudden loss of mass from a moving star.

\noindent In scalar notation, if $\theta$ denotes the angle between the kick vector and the pre-SN orbital velocity of the exploding star, the magnitude of the post-SN systemic velocity is given by:
\begin{equation}
  v_{\rm sys}^2  = \frac{M_{\rm NS}^2\,w^2
                  +(M_{\rm He,f}-M_{\rm NS})^2\,v_{\rm He}^2
                  -2\,M_{\rm NS}\,(M_{\rm He,f}-M_{\rm NS})\,w\,v_{\rm He,f}\,\cos\theta
  }
  {(M_{\rm NS}+M_2)^2}\;,
\end{equation}
and where for a circular pre-SN orbit:
\begin{equation}
  v_{\rm He,f} = \frac{M_2}{M_{\rm He,f}+M_2}\;\sqrt{\frac{G\,(M_{\rm He,f}+M_2)}{a_0}}\;.
\end{equation}


\clearpage
\noindent
{\bf {\large Further observational constraints on the kick in \eDNS}}\\
\noindent
In this work, we have argued that \eDNS originated from an asymmetric SN with a kick, $w\gtrsim 220\;{\rm km\,s}^{-1}$. The main piece of evidence comes from the current spin period (47.5~ms) of the recycled pulsar which places constraints on the last mass-transfer stage. Detailed calculations of this mass transfer (Case~BB Roche-lobe overflow (RLO)) prior to the SN point to an exploding He-star mass of $M_{\rm He,f}\le2.0\;M_\odot$, i.e. considerably smaller than the mass necessary in a symmetric SN ($M_{\rm He,f}=4.19\;M_\odot$) to reproduce the extreme eccentricity observed in this system. 
It is also possible to observationally constrain the kick, and thus discriminate between a symmetric SN and an asymmetric SN, based on $v_{\rm sys}$. If the SN was symmetric, the pre-SN orbital separation ($a_0$) is equal to the post-SN periastron separation $q=a(1-e)$. 
Taking here the value $a_0 \simeq 0.832\;R_\odot$ then implies that the relative velocity between the two components prior to the SN was $v_{\rm rel}=1133\;{\rm km\,s}^{-1}$, and thus the center-of-mass velocity of the exploding He-star was $v_{\rm He,f}=v_{\rm rel}\,(M_2/M_0)=285\;{\rm km\,s}^{-1}$. Finally, conservation of momentum thus yields a post-SN systemic (recoil) velocity of $v_{\rm sys}=v_{\rm He,f}\,(\Delta M/M) = 283\;{\rm km\,s}^{-1}$.
From Fig.~4, we notice from the near one-to-one correlation between $w$ and $v_{\rm sys}$ that such a large value for $v_{\rm sys}$ is only possible for asymmetric SNe if the kick was rather large, i.e. $w\simeq 570\;{\rm km\,s}^{-1}$. On the other hand, if proper motion analysis points to e.g. $v_{\rm sys}\simeq 100\;{\rm km\,s}^{-1}$ then we can constrain $250 < w<300\;{\rm km\,s}^{-1}$.

\vspace{-0.1cm}
A newborn NS with a gravitational mass of $1.40^{+0.05}_{-0.01}\;M_\odot$ corresponds to a baryonic mass of about $1.60\;M_\odot$, depending on the NS equation of state\cite{SI:ly89}. If the SN ejects no material, this also represents the {\em minimum} mass of the collapsing (naked) He star. Therefore, given the $1\sigma$ uncertainties in the mass of the newborn NS companion in \eDNS, we adopt a conservative lower limit of $1.60\;M_\odot$ for the exploding He-star mass in our ``realistic'' simulations.

Ref.~6 presented evidence that the NS kick magnitude increases with the metal core mass of the exploding star, and thus with the mass of the newborn NS (and that an average trend may persist between DNS eccentricity and NS mass, despite large scatter resulting from various kick directions). A large kick and a high eccentricity are therefore natural for \eDNS given its relatively massive NS companion. To produce a large kick from the {\em tug-boat mechanism}\cite{SI:wjm13,SI:gj18} most likely requires SN ejecta of a few $0.1\;M_\odot$, and for this reason we may anticipate that the exploding He-star mass is closer to $2.0\;M_\odot$ than $1.60\;M_\odot$. For comparison, the Hulse--Taylor pulsar\cite{SI:wnt10} (PSR~B1913+16) also has a relatively massive second-born NS ($1.389\;M_\odot$, $e=0.617$), and simulations\cite{SI:tkf+17} show that this NS received a kick of order $w\simeq 200$--$450\;{\rm km\,s^{-1}}$, in good agreement with what we expect for \eDNS.

\clearpage
\noindent
{\bf {\Large B. Marginally bound orbits ($e\rightarrow 1$) and probabilities}}\\
\noindent To determine the probability of obtaining an extremely eccentric post-SN orbit we need to first determine the width of the polar–angle interval ($\Delta\theta$) above the
critical disruption angle, $\theta_{\rm crit}$, for which the post-SN
eccentricity always satisfies $e \ge e_0$, independently of $\phi$. 
We adopt $e_0 = 0.992$, but the derivation is general. (In this section, $e_0$ refers to the {\em current} eccentricity at $t=0$ in Fig.~1).

\noindent
\underline{i) Minimum eccentricity (independent of $\phi$)}\\
The post-SN eccentricity is given by Eq.~(\ref{eq:e-kick}). At fixed $\theta$, the minimum over $\phi$ occurs for $\cos^2\phi=0$:
\begin{equation}
  e_{\min}^2(\theta) = 1 - \xi(2-\xi) \;.
\end{equation}
If this minimum already satisfies $e_{\min}(\theta)\ge e_0$, then $e(\theta,\phi)\ge e_0$ for all $\phi$. Hence, the condition for ``always high eccentricity'' is:
\begin{equation}
  1 - \xi(2-\xi) \;\ge\; e_0^2 \;.
\end{equation}

\noindent
\underline{ii) Condition on $\xi$}\\
Rewriting the above condition yields:
\begin{equation}
  \xi^2 - 2\xi + (1 - e_0^2) \ge 0 \;.
\end{equation}
The roots of the quadratic are: $\xi = 1 \pm e_0\;$, 
so the inequality is satisfied for
\begin{equation}
  \xi \le 1 - e_0 \quad \mbox{or} \quad \xi \ge 1 + e_0\;.
\end{equation}
Near disruption we have $\xi \approx 2$, so the relevant boundary is $\xi = 1 + e_0$.

\noindent
\underline{iii) Convert the $\xi$ boundaries to $\theta$ boundaries}\\
We recall that the critical disruption angle $\theta_{\rm crit}$ is given by Eq.~(\ref{eq:theta_crit}):
\begin{equation}
  \cos\theta_{\rm crit} = \frac{2b - 1 - k^2}{2k}\;.
\end{equation}
The angle $\theta_e$ for which $e_{\min}(\theta_e) = e_0$ therefore satisfies:
\begin{equation}
  \frac{1 + k^2 + 2k\cos\theta_e}{b} = 1 + e_0 \;,
\end{equation}
so:
\begin{equation}
\cos\theta_e = \frac{b(1 + e_0) - (1 + k^2)}{2k} \;.
\end{equation}
Thus the width of the interval above disruption for which $e\ge e_0$ for all $\phi$ is:
\begin{equation}\label{eq:ecc_width}
\Delta\theta = \theta_e - \theta_{\rm crit}
= \cos^{-1}\!\left(\frac{b(1 + e_0) - (1 + k^2)}{2k}\right) - \cos^{-1}\!\left(\frac{2b - 1 - k^2}{2k}\right)\;.
\end{equation}
For the specific numerical values taken from the example in Extended Data Fig.~6 ($P_0=0.05\;{\rm days}$, $M_{\rm He,f}=1.90\;M_\odot$, $M_{\rm NS}=1.40\;M_\odot$, $M_2=1.41\;M_\odot$ and $w=350\;{\rm km\,s}^{-1}$) and taking $e_0=0.992$ for \eDNS, we find $b\simeq 0.8489,\; k\simeq 0.406$, which gives: $\Delta\theta \simeq 0.632^\circ$. The spread in eccentricity with angle $\phi$ is only $\delta e<4.6\times10^{-4}$ over the entire interval $\phi\in [0,\,2\pi [$.

\noindent
\underline{iv) Near-disruption approximation}\\
Because $\Delta\theta$ is small, one may linearise $\xi(\theta)$ about
$\theta_{\rm crit}$. Expanding
\begin{equation}
  \xi(\theta) \simeq 2 + \left.\frac{d\xi}{d\theta}\right|_{\theta_{\rm crit}} \!\!\!(\theta - \theta_{\rm crit}) \;,
\end{equation}
and requiring $\xi(\theta_e) = 1 + e_0$, gives:
\begin{equation}
  \Delta\xi \equiv \xi(\theta_e)-2 = -(1 - e_0) \;.
\end{equation}
Since
\begin{equation}
  \frac{d\xi}{d\theta} = -\frac{2k}{b}\,\sin\theta \;,
\end{equation}
we obtain:
\begin{equation}\label{eq:approx-width}
  \Delta\theta \equiv \theta - \theta_{\rm crit}\simeq \frac{(1 - e_0)\,b}{2k\,\sin\theta_{\rm crit}} \quad {\rm (in~radians)}.
\end{equation}
For our parameters from above, this approximation yields $\Delta\theta \simeq 0.01109~{\rm rad} \simeq 0.635^\circ\;$, in excellent agreement with the exact solution ($\Delta\theta \simeq 0.632^\circ$) from Eq.~(\ref{eq:ecc_width}).

\noindent
\underline{v) Probability for extreme eccentricity post-SN orbit}\\
For an isotropic kick direction, the probability density for the polar angle $\theta \in [0,\,\pi ]$ is:
\begin{equation}
  P(\theta)\,d\theta = \frac{1}{2}\sin \theta\,d\theta\;,
\end{equation}
and the cumulative probability for a near-disruption angle within the band $\theta \in [\theta_{\rm crit},\,\theta_{\rm crit}+\Delta\theta]$ is: 
\begin{equation}\label{eq:prob-disrupt}
  P_{\Delta\theta_{\rm crit}} = \int_{\theta_{\rm crit}}^{\theta_{\rm crit}+\Delta\theta} \frac{1}{2}\sin\theta\,d\theta
= \frac{\cos\theta_{\rm crit} -\cos(\theta_{\rm crit}+\Delta\theta)}{2}\;.
\end{equation}

\noindent
\underline{vi) Near-disruption approximation for the probability}\\
For small $\Delta\theta$, we may expand the cosine in the above equation to first order
around $\theta_{\rm crit}$:
\begin{equation}
  \cos(\theta_{\rm crit} + \Delta\theta)
  \simeq \cos\theta_{\rm crit} - \sin\theta_{\rm crit}\,\Delta\theta\;.
\end{equation}
Inserting this approximation into Eq.~(\ref{eq:prob-disrupt}) we obtain:
\begin{equation}
  P_{\Delta\theta_{\rm crit}} \simeq \frac{\cos\theta_{\rm crit} - \left[\cos\theta_{\rm crit}
  - \sin\theta_{\rm crit}\,\Delta\theta\right]}{2}
  = \frac{\sin\theta_{\rm crit}\,\Delta\theta}{2}\;.
\end{equation}
Using the near-disruption approximation for the angular width (Eq.~\ref{eq:approx-width}), we finally obtain the remarkably simple expression:
\begin{equation}
  P_{\Delta\theta_{\rm crit}}
  \simeq \frac{\sin\theta_{\rm crit}}{2}\,
         \frac{(1 - e_0)\,b}{2k\,\sin\theta_{\rm crit}}
  = \frac{(1 - e_0)\,b}{4k}\;.
\end{equation}
Thus, in the near-disruption limit, the probability that an isotropic kick produces a marginally bound, extremely eccentric post-SN orbit with $e \ge e_0$ is approximately:
\begin{equation}
  P(e \ge e_0) \simeq \frac{(1 - e_0)\,b}{4k}\;,
\end{equation}
independent of $\theta_{\rm crit}$.

\noindent
{\bf {\large Probability of producing a system like \eDNS}}\\
\noindent
For the specific example discussed above (cf. Extended Data Fig.~6 for \eDNS), we had
\begin{equation}
  b \simeq 0.8489,\quad
  k \simeq 0.406,\quad   
  e_0 = 0.992,
\end{equation}
so that
\begin{equation}
  P(e \ge e_0)
  \simeq \frac{(1 - e_0)\,b}{4k}
  = \frac{0.008 \times 0.8489}{4 \times 0.406}
  \simeq 4.2\times 10^{-3}.
\end{equation}
In other words, for an isotropic kick with magnitude $w=350\;{\rm km\,s^{-1}}$ in this particular pre-SN binary, the probability of forming a marginally bound post-SN system with $e \ge 0.992$ is $0.42\%$. Numerical simulations based on $10^8$ SNe yield a consistent value of $0.43\%$ (Extended Data Fig.~7).
Restricting attention to systems that {\em survive} the SN, this fraction increases to $\sim 0.52\%$. For kicks of $w=300\;{\rm km\,s^{-1}}$ and $w=500\;{\rm km\,s^{-1}}$, the corresponding probabilities for the surviving systems are $\sim 0.54\%$ and $\sim 0.49\%$, respectively. Allowing the exploding helium-star mass to vary within $1.70\le M_{\rm He,f}/M_\odot \le 2.20$ and the kick magnitude within $w=300-500\;{\rm km\,s^{-1}}$, the probability that a surviving system has $e\ge 0.992$ ranges from $0.48\%$ to $0.57\%$ for $P_0=0.05\;{\rm days}$.

Although these probabilities for producing \eDNS are small, they are not exceptional. For the same pre-SN configuration, the probability density of producing DNS systems with moderate eccentricities (e.g.\ $e\simeq 0.2$) is comparable: restricting to $e=0.192-0.200$ yields $0.53\%$. This reflects the general requirement that reproducing any specific DNS system --- with given $(e, P)$ --- involves a comparable degree of natal-kick fine-tuning.

An additional illustration can be made using the SN simulations shown in Extended Data Fig.~2: selecting systems with $e=0.490-0.494$ and $P=1.79\;{\rm days}\pm 3\%$ yields an order of magnitude fewer solutions than for \eDNS ($e=0.990-0.994$ and $P=79.1\;{\rm days}\pm 3\%$) under the same setup. Achieving a comparable formation probability for the $e\sim 0.50$ systems requires either relaxing the orbital-period tolerance to $\sim 30\%$ or adopting a different prior on pre-SN orbital periods (e.g.\ flat in $\log P_0$).

Thus, the formation of \eDNS does not require exceptional fine-tuning, but instead reflects the generic requirement that reproducing {\em any} specific DNS system --- with given $(e, P)$ --- demands a comparable degree of natal-kick fine-tuning.

\clearpage
\noindent
{\bf {\Large C. Recycling and amount of mass accreted prior to SN}}\\
\noindent
Considering the spin angular momentum added to an accreting NS the {\em minimum} mass, $\Delta M_{\rm eq}$, required to reach a given recycled spin period is given by\cite{SI:acrs82,SI:tlk12}:
\begin{eqnarray}\label{eq:deltaM}
     \Delta M_{\rm eq} & = & I\;\left( \frac{\Omega_{\rm eq}^4}{G^2M^2} \right) ^{1/3} \,f(\alpha,\xi,\phi,\omega_c) \\ \nonumber 
                       & \simeq & 0.22\,M_{\odot}\; \frac{(M/M_{\odot})^{1/3}}{P_{\rm ms}^{4/3}}
\end{eqnarray}
where $M$ is the NS mass, $P_{\rm ms}$ is the spin period in ms, $I$ is the moment of inertia, $\Omega_ {\rm eq}$ is the NS spin angular velocity, which at equilibrium configuration is equal to the Keplerian angular velocity of matter at the magnetospheric boundary where the accreted matter enters the magnetosphere, and, finally, where  $f(\alpha,\xi,\phi,\omega_c)\simeq1$ is a numerical factor that depends on geometry and disk-magnetosphere interaction details\cite{SI:tv23}. 
Given the current spin period of $P_{\rm spin}=47.5\;{\rm ms}$ and a pulsar mass $M\simeq 1.4\;M_\odot$, we find that \eDNS must have accreted: $\Delta M>1.4\times 10^{-3}\;M_\odot$.

\clearpage
\noindent
{\bf {\Large D. Spin-down since recycling (post-SN spin evolution)}}\\
\noindent
Numerical simulations\cite{SI:tlp15,SI:jtcf21} demonstrate that core collapse of the mass-transferring He-star occurs shortly after (or even during) the final phase of RLO (recycling). The initial post-SN spin period of the recycled pulsar (first-born NS) is thus $P_{\rm spin,SN}=2\pi/\Omega_{\rm eq}$ (see Eq.~\ref{eq:deltaM}).
The spin evolution of a radio pulsar can be computed via integration of the spin-deceleration equation, assuming that the spin-down is characterized by a constant braking index, $n\equiv \Omega\ddot{\Omega}/\dot{\Omega}^2$,
which relates the braking torque ($N=dJ_{\rm spin}/dt=I\,\dot{\Omega}$) to the rotational angular velocity via $\dot{\Omega}\propto -\Omega^{n}$.
For a pulsar evolving with a constant value of $n$, the kinematic solution for the spin period $P$ at time $t$ (positive in the future, negative in the past) is thus given by\cite{SI:tv23}:
\begin{equation}
      P = P_0 \left[ 1 + (n-1) \frac{{\dot P}_0}{P_0}\,t \right]^{1/(n-1)}\;,
  \label{eq:intro-spindown}
\end{equation}
and
\begin{equation}
  \dot P = {\dot P}_0 \left( \frac{P}{P_0} \right)^{2-n}\;,
  \label{eq:intro-spindown_dot}
\end{equation}
where $P_0$ and $\dot{P}_0$ represent values at $t=0$. Similarly, the true age can be written as:
\begin{equation}
      t=\frac{P}{(n-1)\dot{P}} \left[ 1- \left(\frac{P_0}{P}\right)^{n-1} \right]\;.
  \label{eq:trueage}
\end{equation}
Assuming spin-down with a constant braking index in the standard range $2\le n \le 5$, the true age is confined to the interval $ \frac{1}{2}\,\tau \le t \le 2\,\tau$, obtained from the requirement of a birth spin period $P_0>0$. Here, $\tau$ is the characteristic age defined as $\tau \equiv P/(2\dot{P})$.
However, as a result of a short-lived final mass-transfer phase (Case BB RLO), DNS pulsars are usually only mildly\cite{SI:tkf+17} recycled to $P_0\gtrsim 20\;{\rm ms}$. Based on this, and the spin-evolution tracks in the lower-right panel of Fig.~1, we estimate an upper limit on the true age of \eDNS of $t\lesssim \tau = 6.58\;{\rm Gyr}$.

\newpage
\noindent
{\bf {\Large E. Gravitational wave radiation from a highly eccentric DNS}}\\
Calculating GW radiation from \eDNS, we are using a point-mass, quadrupole secular description\cite{SI:pet64}, and safely ignoring effects that are insignificant for a DNS system, like tides, spin-orbit couplings, and other processes --- unless the DNS is drastically influenced by external torques (e.g. in a dense stellar environment like a globular cluster), but this is unlikely for \eDNS given its location in the Galactic field. 

\noindent
{\bf Evolution of the periastron separation}\\
\noindent
The secular (orbit-averaged) evolution of the orbital elements due to GW emission at quadrupole order\footnote{Although the eccentricity is extreme, the orbit remains deeply in the weak-field regime, with a pericenter distance of $q\sim 1.4\times 10^5\;GM/c^2$ and $v_{\rm peri}/c=0.00377$. For comparison, at 3.5PN order $(v_{\rm peri}/c)^{7} \sim 10^{-17}.$} is given by\cite{SI:pet64}\,:
\begin{align}
  \dot{a} \equiv \Big\langle \frac{da}{dt} \Big\rangle  &= -\beta\,\frac{1}{a^3 (1-e^2)^{7/2}} 
                   \left(1+\frac{73}{24}e^2+\frac{37}{96}e^4 \right) \label{eq:adot} \;,\\[1ex]
  \dot{e} \equiv \Big\langle \frac{de}{dt} \Big\rangle  &= -\frac{19}{12}\beta\,\frac{e}{a^4 (1-e^2)^{5/2}}
                   \left(1+\frac{121}{304}e^2 \right)   \label{eq:edot}  \;,
\end{align}
where (for two NSs with gravitational masses $m_1$ and $m_2$):
\begin{equation}
  \beta \equiv \frac{64}{5}\frac{G^3 m_1 m_2 (m_1+m_2)}{c^5}\;.
\end{equation}
The periastron separation is defined as:
\begin{equation}
  q \equiv a(1 - e)\;,
\end{equation}
and thus evolves according to:
\begin{equation}\label{eq:qdot_simple}
  \dot{q} = (1 - e)\,\dot{a} - a\,\dot{e}\;.
\end{equation}

\noindent Substituting the expressions in Eqs.(\ref{eq:adot}) and (\ref{eq:edot}) gives:
\begin{equation}\label{eq:qdot}
\dot{q} = -\frac{G^3 m_1 m_2 M}{c^5}\,\frac{1}{a^3(1 - e^2)^{7/2}}
\Bigg[
\frac{64}{5}(1 - e)\!\left(1 + \frac{73}{24}e^2 + \frac{37}{96}e^4\right)
- \frac{304}{15}(1 - e^2)e\!\left(1 + \frac{121}{304}e^2\right)
\Bigg].
\end{equation}
At very high eccentricities, the two terms inside the square brackets of Eq.~(\ref{eq:qdot}), originating from $(1-e)\dot a$ and $-a\dot e$, respectively, nearly cancel under GW radiation reaction, such that $\dot q$ may change sign and the periastron separation $q$ remains nearly constant, or even increases slightly, during the early inspiral, consistent with Peters' invariant. Only after significant circularization does $q$ begin to decrease appreciably, as summarized in Supplementary Table~1.

\begin{table*}
\centering
\caption*{{\bf Supplementary Table~1.\;\;
Qualitative evolution of the periastron separation
$q\equiv a(1-e)$ under GW radiation reaction in different
eccentricity regimes.}
The sign of $\dot q$ follows from the competition (see
Eq.~\ref{eq:qdot}) between the two terms
$(1-e)\dot a$ and $-a\dot e$ in
$\dot q=(1-e)\dot a-a\dot e$
(Eq.~\ref{eq:qdot_simple}).}
\label{table:q-evolve}

\begin{tabular}{lccp{6.5cm}}
\hline
\hline
Eccentricity regime
& Dominant term
& Sign of $\dot q$
& Physical interpretation \\
\hline

$e\rightarrow1$
&
Near cancellation
&
$\simeq0^{-}$
&
The two contributions
$(1-e)\dot a$
and
$-a\dot e$
almost exactly cancel, producing a nearly
``frozen'' periastron separation over long
timescales. \\[1ex]

Moderate $e$
&
$(1-e)\dot a$
&
$<0$
&
As the orbit circularizes, shrinkage of the
semi-major axis dominates and the periastron
distance decreases monotonically. \\[1ex]

$e\rightarrow0$
&
$q\rightarrow a$
&
$<0$
&
In the circular limit, the periastron separation
equals the orbital radius and decreases
monotonically due to GW-driven inspiral. \\

\hline
\end{tabular}
\end{table*}

\clearpage
\medskip
\noindent
{\bf Why $q$ appears constant at high eccentricity}\\
\noindent
Using the Peters' invariant, $c_0$:
\begin{equation}
c_0 = a\,\frac{(1-e^2)}{e^{12/19}}\,\left(1+\frac{121}{304}e^2\right)^{-870/2299}\;,
\label{eq:peters_invariant}
\end{equation}
the periastron separation becomes:
\begin{equation}
  q(e) \equiv a(1 - e)
      = c_0\,\frac{e^{12/19}}{1 + e} \left(1 + \frac{121}{304}e^2\right)^{870/2299}\;.
\label{eq:peters_e_to_1}
\end{equation}
In the limit $e \to 1$, all eccentricity-dependent factors asymptote to constants, so the periastron separation varies only weakly with $e$ and remains nearly constant until the orbit circularizes significantly. More precisely, in the formal limit $e \to 1$ one finds:
\begin{equation}
  q(e) \simeq 0.566\,c_0\;,
\end{equation}
showing explicitly why the periastron separation remains approximately constant during the early inspiral.
We retain the full eccentricity dependence in Eq.~(\ref{eq:peters_e_to_1}), which increases the $e\to1$ scaling coefficient of $q(e)$ by about 13\% relative to eq.~(5.12) of Peters~(1964).

\medskip
\noindent
{\bf Numerical example}\\
\noindent
For a DNS (\eDNS) with $m_1 = 1.41\;M_\odot$, $m_2 = 1.40\;M_\odot$, $P_{\rm orb} = 79.1$~days ($a_0 \simeq 109.4\;R_\odot$), and $e_0 = 0.9924$, the current periastron distance is:
\begin{equation}
  q_0 = a_0(1 - e_0) \approx 0.8316\;R_\odot \;.
\end{equation}
Backward integration (using the Peters' invariant) to $t = -13.8$~Gyr yields
$a \simeq 3321\;R_\odot$, $e \simeq 0.99975$, and
\begin{equation}
  q_{\rm Hubble} \simeq 0.8338\;R_\odot \;.
\end{equation}
Thus over a Hubble time, $q$ decreases by only $\sim 2.1\times10^{-3}\;R_\odot$ ($\simeq 1480$~km), demonstrating the ``nearly frozen'' pericentre regime; 
see also Supplementary Table~1.

\noindent At the present epoch, the rates are:
\begin{equation}
  \dot{a} \simeq -7.21\times10^{-8}\,R_\odot\,{\rm yr^{-1}}, \qquad
  \dot{e} \simeq -4.99\times10^{-12}\,{\rm yr^{-1}}, \qquad
  \dot{q} \simeq -1.45\times10^{-12}\,R_\odot\,{\rm yr^{-1}},
\end{equation}
i.e.\ the periastron is shrinking by only $\sim 10^{-12}\,R_\odot$ per year while $a$ decreases much faster. We note that although $q$ has increased slightly over most of the system's lifetime, it is currently decreasing as the orbit continues to circularize.

\medskip
\noindent
\underline{Summary of the periastron evolution}\\
\noindent At very high eccentricities, $q$ is approximately constant because GW losses remove angular momentum most efficiently near periastron, causing the apastron to collapse while the periastron shrinks only slowly (Fig.~2, lower-left panel). Once the system circularizes, however, $q$ begins to decrease significantly, approaching zero at merger.

\noindent
{\bf {\large GW luminosity and burstiness at the current separation}}\\
In the following we calculate the GW luminosity as a function of orbital phase (true anomaly), demonstrating that it varies by more than nine orders of magnitude over the orbit.

\noindent
{\bf Orbit-averaged GW luminosity and energy radiated per orbit by \eDNS}\\
The orbit-averaged GW luminosity and the corresponding energy radiated per orbit are given by\cite{SI:pm63}:
\noindent
\begin{equation}\label{eq:Lavg}
  \langle L\rangle = \frac{32}{5}\,\frac{G^4}{c^5}\,\frac{m_1^2 m_2^2 (m_1+m_2)}{a^5}\,
                     \frac{1+\frac{73}{24}e^2+\frac{37}{96}e^4}{(1-e^2)^{7/2}}\;.
\end{equation}
For $m_1=1.41\;M_\odot$, $m_2=1.40\;M_\odot$, $P=79.1\;{\rm days}$ ($a=109.4\;R_\odot$), $e=0.9924$:
\begin{equation}
  \langle L\rangle \;=\; 7.14\times 10^{29}\;{\rm erg\,s}^{-1}\; \approx\; 1.87\times 10^{-4}\,L_\odot .
\end{equation}
The energy radiated per orbit is:
\begin{equation}\label{eq:Eorbit}
  \Delta E_{\rm GW} \;=\; \langle L\rangle\,P\;=\; 4.88\times 10^{36}\;{\rm erg}.
\end{equation}

\noindent
{\bf Instantaneous ``circular-equivalent'' luminosity at periastron and apastron (proxy)}\\
\noindent
Let $\mu = m_1 m_2/(m_1+m_2)$ and define the GW luminosity of a circular binary with instantaneous separation $r$ as:
\begin{equation}\label{eq:Lcirc}
  L_{\rm circ}(r) = \frac{32}{5}\,\frac{G^4}{c^5}\,\frac{\mu^2 (m_1+m_2)^3}{r^5}\;.
\end{equation}
Evaluating this expression at the periastron and apastron separations, $r_{\rm peri}=q=a(1-e)$ and $r_{\rm ap}=a(1+e)$, respectively, yields:
\begin{equation}
  L_{\rm circ}(r_{\rm peri}) = 2.75\times 10^{33}\ {\rm erg\,s^{-1}} \;\simeq\; 0.72\,L_\odot,\quad 
  L_{\rm circ}(r_{\rm ap}) = 2.22\times 10^{21}\ {\rm erg\,s^{-1}} \;\simeq\; 5.8\times 10^{-13}\,L_\odot.
\end{equation}
In simple terms, the luminosity contrast scales as $L_{\rm GW,peri}/L_{\rm GW,ap}=[(1+e)/(1-e)]^5 \simeq 10^{12}$, illustrating the strongly periastron-peaked (``bursty'') nature of the GW emission.

\begin{table*}[h]
\centering
\caption*{{\bf Supplementary Table~2.\;\;GW burstiness near periastron.}
Despite spending only a fraction $\lesssim10^{-3}$ of each orbit within a few times the periastron separation ($q$), the binary radiates $\gtrsim 99\%$ of its GW energy there (for $e=0.9924$). This highlights the extreme periastron dominance of GW emission from highly eccentric DNS systems. Here $\nu_k$ is defined by $r(\nu_k)=kq$.}
\label{fig:periastron_energy_fraction}
\begin{tabular}{ccccc}
\hline\hline
$k$ & Region & Time fraction $f_k$ & Energy fraction $E_{(r<k\cdot q)}/\Delta E_{\rm GW}$ & $|\nu_k|$ [deg] \\
\hline
$2$ & $r<2\;q$ & $3.99\times10^{-4}$ & $98.33\%$   & $90.22^\circ$ \\
$3$ & $r<3\;q$ & $7.07\times10^{-4}$ & $99.73\%$   & $109.78^\circ$ \\
$5$ & $r<5\;q$ & $1.40\times10^{-3}$ & $99.97\%$   & $127.31^\circ$ \\
\hline
\end{tabular}
\end{table*}

\noindent
{\bf Duty cycle near periastron}\\
\noindent
For a Keplerian binary on an eccentric orbit, the orbital separation $r$ as a
function of true anomaly $\nu$ is given by the standard relation (with $\nu=0$ at periastron):
\begin{equation}\label{eq:rnu}
  r(\nu)=\frac{a(1-e^2)}{1+e\cos\nu}\;.
\end{equation}
The time spent at a given orbital phase follows from conservation of the
specific orbital angular momentum,
$h=r^2\dot{\nu}={\rm const.}$
(or $h=\sqrt{G(m_1+m_2)a(1-e^2)}$), which implies the time–true-anomaly Jacobian reuqired for time-averaging or phase integrals:
\begin{equation}\label{eq:dtdnu}
  \frac{dt}{d\nu} = \frac{r^2}{h} =
  \frac{a^{3/2}(1-e^2)^{3/2}}{\sqrt{G(m_1+m_2)}} \frac{1}{\big(1+e\cos\nu\big)^2}  = \frac{(1-e^2)^{3/2}}{\Omega}\,\frac{1}{\big(1+e\cos\nu\big)^2}\;.
\end{equation}
The prefactor is constant for a given orbit and does not depend on the true anomaly $\nu$; it therefore cancels when computing fractions of the orbital period and can be omitted in what follows, leaving only the $\nu$-dependent scaling:
\begin{equation}
  \frac{dt}{d\nu}\propto \frac{1}{\big(1+e\cos\nu\big)^2}\;.
\end{equation}
Using these relations, the fraction of orbital time spent within a multiple $k$ of the periastron distance ($r<k\,r_{\rm p}$) can be written as:
\begin{equation}
  f_k \;=\; \frac{\displaystyle \int_{\{r<k r_{\rm p}\}} \frac{d\nu}{(1+e\cos\nu)^2}}
                 {\displaystyle \int_{0}^{2\pi} \frac{d\nu}{(1+e\cos\nu)^2}} \;.
\end{equation}
For $e=0.9924$ as in \eDNS, the numerical values are (Supplementary Table~2):
\begin{equation}
  f_{k=2} \;=\; 3.99\times 10^{-4},\quad
  f_{k=3} \;=\; 7.07\times 10^{-4},\quad
  f_{k=5} \;=\; 1.40\times 10^{-3},
\end{equation}
i.e.\ most of $\Delta E_{\rm GW}$ is emitted within a tiny ($\sim10^{-3}$) fraction of the orbital phase surrounding periastron (see Supplementary Fig.~1).
The strong periastron dominance follows because the instantaneous luminosity is so sharply peaked near $\nu=0$ that it more than compensates for the short time spent near periastron.

\begin{figure*}[h]
\vspace*{0.0cm}
\hspace*{1.0cm}
\includegraphics[width=0.9\textwidth]{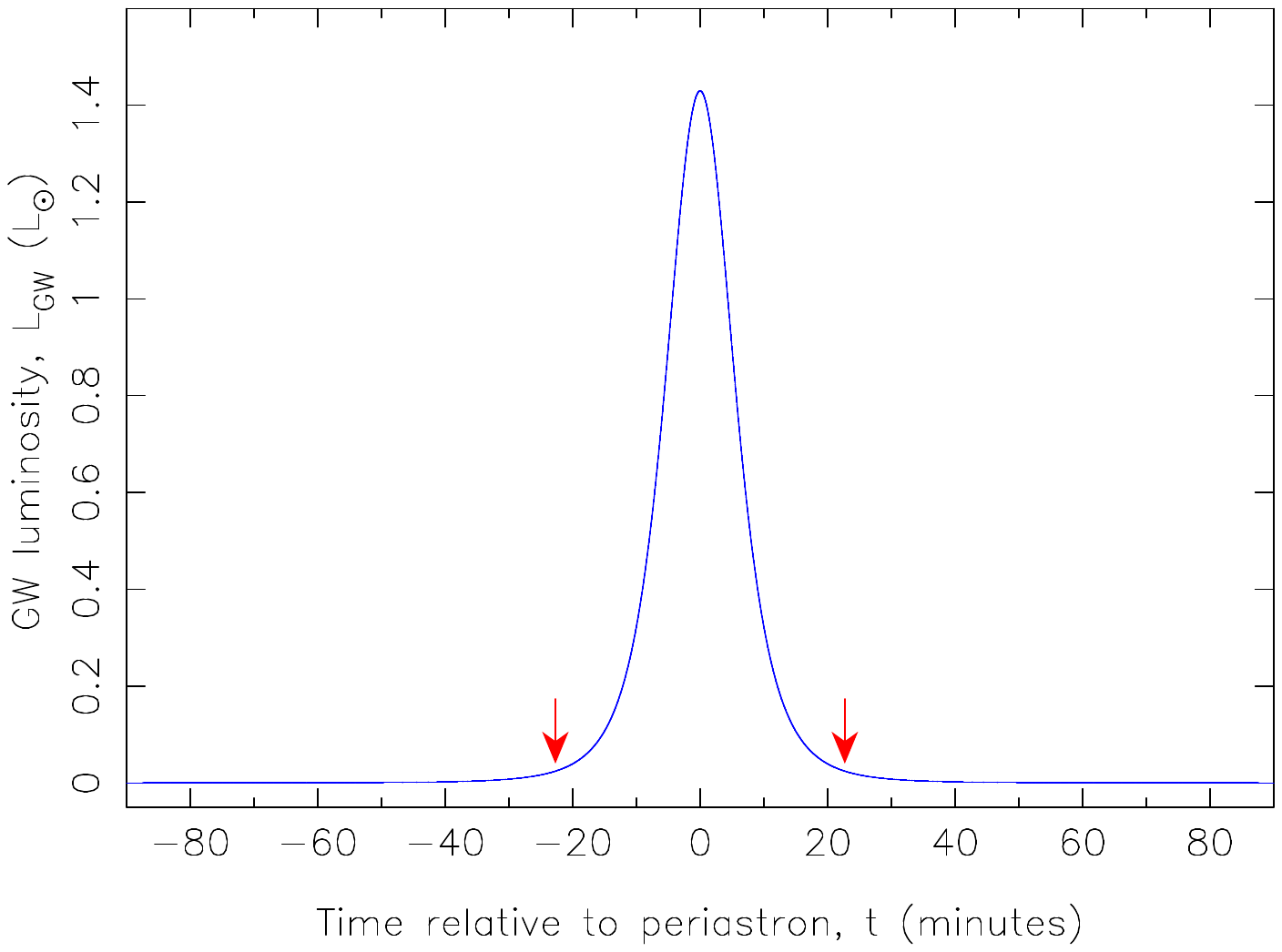}
\caption*{{\bf Supplementary Fig.\,1}\;\;
\textbf{GW luminosity as a function of time near periastron passage for the present-day  \eDNS.} Arrows indicate the times ($t=\pm 22.6\;{\rm min}$) corresponding to true anomalies $\nu=\pm\pi/2$, measured relative to periastron ($t=0$, $\nu=0$), demonstrating that the GW emission is concentrated within a narrow time interval compared to $P_{\rm orb}=79.1\;{\rm days}$.
\label{fig:L_GW}
}
\end{figure*}

\newpage

\noindent
The strong periastron dominance can also be understood from the phase dependence of the instantaneous GW luminosity. 
Given the expression for $r({\nu})$ in Eq.~(\ref{eq:rnu}), the circular-equivalent scaling $L(\nu)\propto r(\nu)^{-5}$ implies $L(\nu)\propto (1+e\cos\nu)^5$, while $dt/d\nu\propto (1+e\cos\nu)^{-2}$. Hence, in this simple proxy:
\begin{equation}
  \frac{dE}{d\nu}=L(\nu)\frac{dt}{d\nu}\;\propto\;\frac{(1+e\cos\nu)^3}{a^{7/2}(1-e^2)^{7/2}}\;.
\end{equation}
Using the full instantaneous quadrupole expression (Eq.~\ref{eq:Lnu}) gives the same qualitative result, with:
\begin{equation}
  \frac{dE}{d\nu}\;\propto\;\frac{(1+e\cos\nu)^2}{a^{7/2}(1-e^2)^{7/2}} \left[12+24e\cos\nu+e^2(1+11\cos^2\nu)\right]\;.
\end{equation}
Thus, the GW energy loss is strongly concentrated toward periastron. For the present system ($e=0.9924$), the numerical integrations in Supplementary~Table~2 show that more than $\sim 99\%$ of the emitted GW energy is released while the stars are within only a few times the periastron distances of each other. Consequently, even though the binary spends only a tiny fraction ($\sim 10^{-3}$) of its orbital period near periastron, this short interval accounts for essentially all of the GW energy loss.

\noindent The true anomaly, $\nu$ changes from $-\pi/2$ to $+\pi/2$ ($-90^\circ$ to $+90^\circ)$ when $t=-22.6\;{\rm min}$ to $t=+22.6\;{\rm min}$, where $t=0$ marks the periastron time.

\noindent
{\bf {\large Phase-resolved (instantaneous) GW luminosity along an eccentric orbit}}\\
The orbit-averaged GW luminosity obscures the fact that, for highly eccentric binaries, the emission is extremely concentrated near periastron. We therefore return to the phase-resolved (instantaneous) GW luminosity along the orbit, which directly reveals the origin of the bursty emission and its relation to the duty cycle.

\noindent
{\bf Instantaneous luminosity (quadrupole order)}\\
\noindent{The instantaneous GW luminosity as a function of true anomaly is\cite{SI:ll75}:
\begin{equation}\label{eq:Lnu}
  L(\nu) \;=\;\frac{8}{15}\,\frac{G^3}{c^5}\,\frac{\mu^2 M^2}{r(\nu)^4}\,
           \Big[\,12\,v(\nu)^2 \;-\; 11\,\dot r(\nu)^2\,\Big]\;,
\end{equation}
where $M=m_1+m_2$ and $\mu=m_1 m_2/M$. 
Equation~(\ref{eq:Lnu}) is obtained by inserting the Newtonian Keplerian motion of two point masses into the standard quadrupole GW luminosity expression $L=\frac{G}{5c^5}\langle\dddot{Q}_{ij}\dddot{Q}_{ij}\rangle$, where $Q_{ij}=\mu\,x_i x_j$ is the mass–quadrupole tensor\cite{SI:pm63,SI:ll75}. It time-averages to the orbit-averaged classical Peters–Mathews luminosity given in Eq.~(\ref{eq:Lavg}).

\noindent
{\bf Keplerian parametrization (true anomaly)}\\
\noindent 
For a Kepler ellipse with semimajor axis $a$ and eccentricity $e$, the relevant relations are:
\begin{equation}
   r(\nu) \;=\; \frac{a(1-e^2)}{1+e\cos\nu}\;,
          \qquad
   \Omega \;=\; \sqrt{\frac{GM}{a^3}}\;,
\end{equation}
\begin{equation}
\label{eq:rdot}
   \dot r(\nu) \;=\; \frac{\sqrt{GM/a}}{\sqrt{1-e^2}}\; e \sin\nu\;,
       \qquad
    r(\nu)\,\dot\phi(\nu) \;=\; \frac{\sqrt{GM a(1-e^2)}}{r(\nu)}\;,
\end{equation}
\begin{equation}
\label{eq:v2}
     v(\nu)^2 \;=\; \dot r(\nu)^2 + \big[r(\nu)\dot\phi(\nu)\big]^2
              \;=\; GM\left(\frac{2}{r(\nu)} - \frac{1}{a}\right) \;.
\end{equation}
Here $h=|\mathbf{r}\times\mathbf{v}|=\sqrt{GMa(1-e^2)}$ is again the constant specific angular momentum of the binary. 
We define the constant mean orbital angular frequency, $\Omega \equiv \sqrt{GM/a^3} = 2\pi/P$ (sometimes denoted $n$ in celestial mechanics) to avoid confusion with the harmonic index, $n$ appearing in Fourier expansions of eccentric GW signals (see below). Notice that $\Omega$ is distinguished from the instantaneous orbital angular velocity $\dot\phi(\nu)=h/r(\nu)^2$, which varies strongly around an eccentric orbit.

\noindent Supplementary Fig.~2 displays the normalized GW energy loss per unit true anomaly, $dE/d\nu$.

To derive the expression for $\dot{r}(\nu)$ given above, we again consider a Kelerian orbit with semimajor axis $a$ and eccentricity $e$. The radial separation as a function of true anomaly $\nu$ is given by Eq.~(\ref{eq:rnu}):
\begin{equation}
  r(\nu) = \frac{a(1-e^2)}{1+e\cos\nu}\;.
\end{equation}
Differentiating with respect to $\nu$ yields:
\begin{equation}
  \frac{dr}{d\nu}= a(1-e^2)\frac{e\sin\nu}{\big(1+e\cos\nu\big)^2}\;.
\end{equation}
For planar Keplerian motion, the specific angular momentum $h \equiv r^2\dot\phi$ is conserved, and the true anomaly advances at the rate $\dot\nu = \dot\phi = h/r^2$.

\noindent Using the chain rule:
\begin{equation}
  \dot r = \frac{dr}{d\nu}\frac{d\nu}{dt} \;,
\end{equation}
and substituting for $d\nu/dt$ and inserting $r(\nu)$ gives: 
\begin{equation}
  \dot r = a(1-e^2)\frac{e\sin\nu}{\big(1+e\cos\nu\big)^2}\;\frac{h}{a^2(1-e^2)^2/(1+e\cos\nu)^2} = \frac{h}{a(1-e^2)}\,e\sin\nu\;.
\end{equation}
Finally, expressing the specific angular momentum as $h = \sqrt{GMa(1-e^2)}$ one obtains the expression for $\dot{r}(\nu)$ in Eq.~(\ref{eq:rdot}).

\begin{figure*}
\centering
\includegraphics[width=0.9\linewidth]{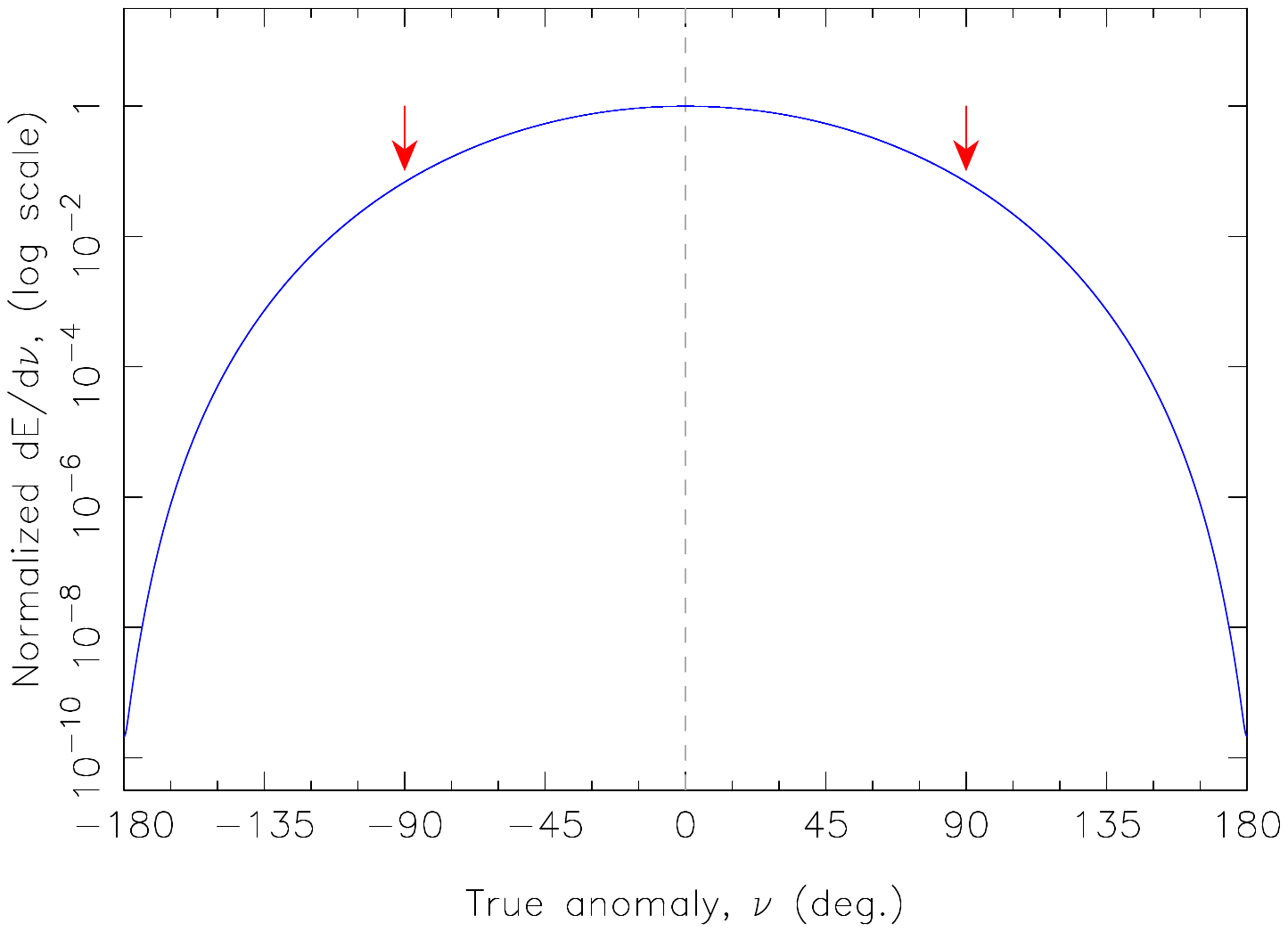}
\\
\vspace{0.2cm}
\caption*{{\bf Supplementary Fig.\,2\;\;Normalized GW energy loss per unit true anomaly, $dE/d\nu$.} Plot combining Eqs.~(\ref{eq:Lnu}) and (\ref{eq:dtdnu}) for the present-day \eDNS system ($m_1=1.41\;M_\odot$, $m_2=1.40\;M_\odot$, $e=0.9924$, $a=109.4\;R_\odot$). The sharp peak at $\nu=0^\circ$ illustrates the extremely periastron-dominated (``bursty'') GW emission: about nine orders of magnitude more energy is radiated near periastron than near apastron. See also Supplementary Fig.~1.}
\label{fig:dEdnu}
\end{figure*}

\clearpage
\noindent
{\bf {\large Frequency band of emitted GWs}}\\
For a circular binary with $P_{\rm orb}=79.1\;{\rm days}$, the dominant GW frequency would be $f_{\rm GW}\simeq 2/P_{\rm orb}\simeq 2.9\times10^{-7}\;{\rm Hz}$, far below the LISA\cite{SI:LISA+23} and TianQin bands\cite{SI:llt+25}. However, for an eccentric system the emission is distributed over harmonics $f_n=n/P_{\rm orb}$ (the circular limit is dominated by $n=2$). A simple estimate of the harmonic number corresponding to the strongest emission is:
\begin{equation}
  n_{\rm peak}\sim \frac{\Omega_{\rm peri}}{\Omega} = \frac{(1+e)^{1/2}}{(1-e)^{3/2}} \;,
\end{equation}
where $\Omega_{\rm peri}\equiv v_{\rm peri}/q$ is the instantaneous angular frequency at pericentre and from Eq.~(\ref{eq:v2}) we have $v_{\rm peri}=\sqrt{GM\left(2/q - 1/a\right)}$. For $e=0.9924$, this estimate yields $n_{\rm peak}\simeq2.1\times10^3$. A more accurate fit (the Wen approximation\cite{SI:wen03}) to the Peters--Mathews harmonic spectrum function gives:
\begin{equation}\label{eq:Wen03}
   n_{\rm peak} \simeq \frac{2\,(1+e)^{1.1954}}{(1-e^2)^{3/2}}\simeq 2447 \;.
\end{equation}
Direct evaluation of the $g(n,e)$ function\cite{SI:pm63} from Eq.~(\ref{eq:PM_gne}) places the maximum at $n_{\rm peak}=2468$, confirming the accuracy of this approximation (see also Supplementary Fig.~3). Hence, the bursty GW emission from our DNS system peaks at:
\begin{equation}
  f_{\rm peak,GW} = \frac{n_{\rm peak}}{P_{\rm orb}} \simeq 3.6\times10^{-4}\;{\rm Hz}\simeq 0.36\;{\rm mHz}\;,
\end{equation}
which lies within the low-frequency part of the LISA and TianQin sensitivity bands.

\noindent
{\bf {\large Detectability and inspiral}}\\
For a highly eccentric binary, the GW power is distributed over many harmonics of the orbital frequency rather than being concentrated at a single frequency \cite{SI:kcb+18,SI:ds18,SI:abp+20}. The emission is strongest near the dominant harmonic, but the strain in any individual harmonic is reduced compared to that of a circular binary radiating the same total power at a single frequency. Hence, estimates based only on the circular-binary characteristic strain evaluated at $f_{\rm peak,GW}$ can overestimate the observable signal. We therefore estimate the contribution from each harmonic using the Peters--Mathews power distribution and sum the corresponding signal-to-noise contributions in quadrature.

\noindent The $n$-th harmonic is emitted at:
\begin{equation}
  f_n = n f_{\rm orb}\;.
\end{equation}
For a circular binary emitting at GW frequency ($f=2/P_{\rm orb}$), the sky- and polarization-averaged strain amplitude is:
\begin{equation}
   h_{0}=\sqrt{\frac{32}{5}}\,\frac{(G\mathcal{M})^{5/3}}{c^4 d}\,(\pi \,f)^{2/3}\;.
\end{equation}
For an eccentric binary, the strain amplitude in the $n$-th harmonic can be written approximately as:
\begin{equation}
  h_n = h_{0}\,H_n(e)\;,
\end{equation}
where
\begin{equation}
  H_n(e)=\sqrt{g(n,e)}\left(\frac{2}{n}\right)^{5/3}\;.
\end{equation}
Here $g(n,e)$ is the dimensionless Peters--Mathews harmonic power function\cite{SI:pm63}:
\begin{equation}\label{eq:PM_gne}
\begin{split}
   g(n,e) = \frac{n^4}{32}\; \Bigg\{
    &   \Big[ J_{n-2}(ne) -2e\,J_{n-1}(ne) +\frac{2}{n}J_n(ne) +\,2e\,J_{n+1}(ne)  -J_{n+2}(ne) \Big] ^2   \\[0.1cm]
    +& (1-e^2)   \Big[ J_{n-2}(ne) -2J_n(ne)    +J_{n+2}(ne)    \Big]^2  + \frac{4}{3n^2}  \,J_n^2(ne) \Bigg\} \;,
\end{split}
\end{equation}
where $J_k(x)$ denotes the Bessel function of the first kind of integer order $k$, evaluated at argument $x$ (i.e. here $x=ne$, the product of the harmonic number $n$ and eccentricity $e$).
This normalization satisfies:
\begin{equation}
  \sum_{n=1}^{\infty} g(n,e) =
  \frac{ 1+\frac{73}{24}e^2+\frac{37}{96}e^4}{(1-e^2)^{7/2}} \equiv f(e) \;.
\end{equation}
This prescription gives $H_2(0)=1$, thereby recovering the circular limit. The factor $(2/n)^{5/3}$ accounts for the frequency dependence of the GW strain amplitude, $h\propto f^{2/3}$, when comparing the $n$-th harmonic with the quadrupole ($n=2$) harmonic of a circular binary.

\noindent The corresponding characteristic strain in the $n$-th harmonic over an observing time $T_{\rm obs}$ is:
\begin{equation}
h_{c,n}=h_n\sqrt{f_nT_{\rm obs}}\;.
\end{equation}
The matched-filter signal-to-noise ratio (S/N), denoted by $\rho$, follows from the standard expression\cite{SI:mag08}:
\begin{equation}
\rho^2 = 4\int_0^\infty \frac{|\tilde h(f)|^2}{S_n(f)}\,df\;,
\end{equation}
where $S_n(f)$ is the one-sided noise power spectral density. For a quasi-stationary (slowly evolving) eccentric binary, the signal can be approximated as a sum over discrete harmonics and the S/N ($\rho$) is then estimated by summing the contribution from each harmonic:
\begin{equation}
\rho^2\simeq  \sum_n   \frac{h_{c,n}^2}{f_n\,S_n(f_n)}\;.
\end{equation}

For the present system ($e=0.9924$), the Peters--Mathews spectrum peaks near $n_{\rm peak}\simeq 2468$, corresponding to $f_{\rm peak,GW}\simeq 0.36\;{\rm mHz}$. At this harmonic, $H_n(e)\simeq {\rm few}\times10^{-4}$. 
The adopted LISA sensitivity curve, including the detector noise PSD and Galactic confusion noise, follows the formulation of Robson~et~al.\cite{SI:rcl19}. The present-day eccentric signal is therefore far weaker than a circular-binary estimate at the same frequency. A harmonic-summed estimate gives $\rho\sim\mathcal{O} (0.1)$ for an assumed nearby distance of $d=1.0\;{\rm kpc}$ and $T_{\rm obs}=4\;{\rm yr}$ (Extended Data Fig.~9). The assumed distance of $1.0\;{\rm kpc}$ is adopted from ref.~30 and based on the pulsar dispersion measure and two different Galactic electron-density models in the direction of \eDNS.

\noindent In summary, \eDNS will therefore not be detectable by LISA. Related Galactic extremely eccentric DNS systems may, however, become detectable during the late stages of their inspiral when the harmonics shift to higher frequencies and the strain increases\cite{SI:abp+20}.

Using the Peters equations for GW-driven evolution\cite{SI:pet64}, the present binary parameters ($P_{\rm orb}=79.1\;{\rm days}$, $e=0.9924$, $m_1=1.41\;M_\odot$, $m_2=1.40\;M_\odot$) imply a remaining inspiral time of $t_{\rm merge} \simeq 2.54\;{\rm Gyr}$. 
Over the first $\sim 2\;{\rm Gyr}$ of this evolution, while the semimajor axis decreases significantly to $a\simeq 12\;R_\odot$ ($P_{\rm orb}\simeq 2.8\;{\rm d}$ and $f_{\rm orb}=4.1\times 10^{-6}\;{\rm Hz}$), the eccentricity remains very high and only decreases moderately to $e\simeq 0.93$ (Fig.~2 and Extended Data Fig.~9). Consequently, after 2~Gyr, the peak harmonic decreases to $n_{\rm peak}\simeq 90$, which compensates for the increased orbital frequency. The frequency of the dominant GW harmonic therefore remains nearly unchanged, $f_{\rm peak,GW}\simeq 0.37\;{\rm mHz}$, so the system remains effectively stationary within the low-frequency end of the LISA band over cosmological timescales, despite the strong bursty emission near periastron.

\noindent The interesting property that $f_{\rm peak,GW}$ remains more or less constant for the major part of the orbital evolution of a highly eccentric binary ($e\rightarrow 1$) can easily be understood. Using the Peters invariant (Eq.~\ref{eq:peters_invariant}), one finds that for highly eccentric binaries $a \propto (1-e)^{-1}$, which implies
$f_{\rm peak,GW} \propto (1+e)^{1/2} \approx {\rm const}$. Hence, despite a dramatic shrinkage of the orbital period, the peak GW frequency remains nearly unchanged throughout most of the inspiral.

\noindent
{\bf {\large Residual eccentricity at merger}}\\
Upon entering the ground-based detector band, if we define 10~Hz by the quadrupole $n=2$ harmonic (the usual LVK convention: $f_{\rm GW}=2\,f_{\rm orb}$) then at $f_{\rm GW}=10\;{\rm Hz}$: $a\simeq 0.00104\;R_\odot$ (723~km) and $e\simeq 1.0\times 10^{-5}$. If instead we define 10~Hz by the peak GW harmonic using Eq.~(\ref{eq:Wen03}), we obtain essentially the same values ($a\simeq 723\;{\rm km}$ and $e\simeq 1.0\times 10^{-5}$). 
Hence, the remaining residual eccentricity of a progenitor system similar to \eDNS\ would not be detected by LVK\cite{SI:ssm+26}, nor by ET\cite{SI:ET+26}. Regardless of whether the detector-band entry is defined via the quadrupole or peak GW harmonic, the system remains essentially circular ($e\sim10^{-5}$), well below the detectability threshold.
\\
\\

\begin{figure*}[h]
\vspace*{-0.1cm}
\hspace*{0.2cm}
\includegraphics[width=0.90\linewidth]{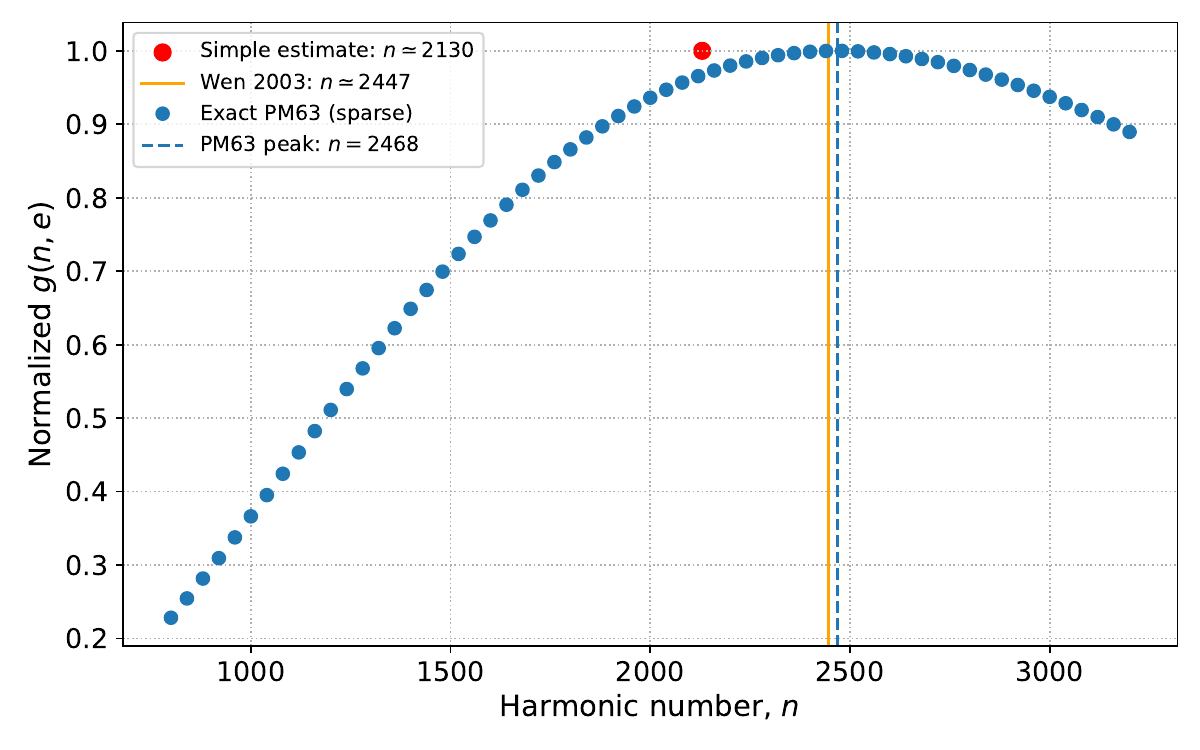}
\caption*{{\bf Supplementary Fig.\,3}\;\;
\textbf{Harmonic power distribution for $e=0.9924$.}
The discrete points show the exact Peters--Mathews function $g(n,e)$ from Eq.~(\ref{eq:PM_gne}), normalised to its maximum value and evaluated on a sparse grid of harmonic indices $n$ around the maximum. Because the distribution is broad ($\Delta n\sim 10^3$), the precise location of the maximum is not important. Different approximations to the peak harmonic --- the simple asymptotic estimate (red point), the Wen approximation (solid orange line), and the exact numerical maximum of the Peters--Mathews function (blue dashed line) --- therefore yield values in the range $n_{\rm peak}\simeq2.1$--$2.5\times10^3$. This variation is physically unimportant: the GW power is distributed over a broad range of harmonics clustered around $n\sim{\rm few}\times10^3$, and the LISA noise PSD varies only slowly across the corresponding frequency interval spanned by the dominant harmonics. For illustrative purposes, the dominant harmonic provides a useful representative frequency, $f_{\rm peak}=n_{\rm peak}f_{\rm orb}$. A quantitative estimate of detectability, however, requires summing the contributions from all harmonics.
\label{fig:harmonics}
}
\end{figure*}

\newpage
\noindent
{\bf {\Large F. Population of known Galactic DNS systems}}\\
There are currently 37 Galactic DNS candidate systems (30 in the Galactic field and 7 in Globular clusters, GCs; Supplementary Table~3). The vast majority are confirmed DNSs. However, systems in GCs or those detected via their young, second-born pulsar often remain unconfirmed.
Data are taken from the ATNF Pulsar Catalogue\cite{SI:mhth05} (version 2.8.0, June~2026) and ref.~30, except for PSR~J1835$-$3259A, which is excluded because it is unlikely to be a DNS system. The recently discovered\cite{SI:lpz+25,SI:yhw+26,SI:yhs+26} DNS systems J1953+1847D, J0641+0448 and J1856$-$0039 are also included.

\begin{table*}[h]
\vspace{-0.5cm}
\centering
\caption*{{\bf Supplementary Table~3.\;\;Known and candidate Galactic DNS systems.} The ten columns list pulsar name, type, spin period, spin-period derivative, orbital period, eccentricity, mass function, total mass, pulsar mass and companion mass, respectively. See text for further details.}
\label{table:DNS_data}

{\footnotesize
 {\begin{tabular}{l c c c c c c c c c}
   \hline
   \hline
\noalign{\smallskip}
\noalign{\smallskip} 
    Pulsar           & Type     & $P_{\rm spin}$  & $\dot{P}_{\rm spin}$ & $P_{\rm orb}$ & $e$ & $f$ & $M_{\rm total}$ & $M_{\rm psr}$ & $M_{\rm c}$\\
                     &          & (ms)   & ($10^{-18}$) & (days)        &       & ($M_\odot$) & ($M_\odot$)     & ($M_\odot$)   & ($M_\odot$) \\ 
   \hline
    J0453+1559       & recycled & 45.8  & 0.186        & 4.07          & 0.113 & 0.196       & 2.73            & 1.56          & 1.17 \\
    J0509+3801       & recycled & 76.5  & 7.93         & 0.380         & 0.586 & 0.0644      & 2.81            & 1.34          & 1.46 \\
    J0528+3529       & recycled & 78.2  & 0.736        & 11.7          & 0.290 & 0.243       & 2.59            & $[1.3]$       & $[1.3]$ \\
    J0641+0448       & recycled & 25.7  & 0.0656       & 3.73          & 0.145 & 0.290       & 2.59            & 1.32          & 1.27 \\
    J0653+0443       & recycled & 47.5  & 0.114        & 79.1          & 0.992 & 0.340       & 2.81            & 1.41          & 1.40 \\
    J0737--3039A     & recycled & 22.7  & 1.76         & 0.102         & 0.0878& 0.291       & 2.59            & 1.34          & 1.25 \\
    J0737--3039B     & young    & 2770  & 892          & $-||-$        & $-||-$& 0.359       & $-||-$          & 1.25          & 1.34 \\
    J1018--1523      & recycled & 83.2  & 0.109        & 8.98          & 0.228 & 0.238       & 2.3             & --            & $>1.16$ \\
    J1155--6529      & recycled & 78.9  & 0.35         & 3.67          & 0.26  & 0.288       & --              & --            & $>1.27$ \\
    J1208--5936      & recycled & 28.7  & $<0.04$      & 0.632         & 0.348 & 0.208       & 2.59            &  $[1.3]$      & $[1.3]$ \\
    J1325--6253      & recycled & 29.0  & 0.048        & 1.82          & 0.0640& 0.142       & 2.57            & $< 1.59$      & $> 0.98$ \\
    J1411+2551       & recycled & 62.5  & 0.0960       & 2.62          & 0.170 & 0.123       & 2.54            & $< 1.62$      & $> 0.92$ \\
    J1518+4904       & recycled & 40.9  & 0.0272       & 8.63          & 0.249 & 0.116       & 2.72            & $< 1.42$      & $>1.29$ \\
    B1534+12         & recycled & 37.9  & 2.42         & 0.421         & 0.274 & 0.315       & 2.68            & 1.33          & 1.35 \\
    J1753--2240      & recycled & 95.1  & 0.970        & 13.6          & 0.304 & 0.0344      & --              & --            & -- \\
    J1755--2550      & young$^*$& 315   & 2430         & 9.70          & 0.0894& 0.0212      & --              & --            & -- \\
    J1756--2251      & recycled & 28.5  & 1.02         & 0.320         & 0.181 & 0.221       & 2.57            & 1.34          & 1.23 \\
    J1757--1854      & recycled & 21.5  & 2.63         & 0.184         & 0.606 & 0.358       & 2.73            & 1.34          & 1.39 \\
    J1759+5036       & recycled & 176   & 0.243        & 2.04          & 0.308 & 0.0819      & 2.62            & $< 1.79$      & $> 0.84$ \\
    J1811--1736      & recycled & 104   & 0.901        & 18.8          & 0.828 & 0.128       & 2.57            & $<1.74$       & $> 0.93$ \\
    J1829+2456       & recycled & 41.0  & 0.0525       & 1.18          & 0.139 & 0.295       & 2.61            & 1.31          & 1.30 \\
    J1844--0128      & recycled & 29.1  & 0.0306       & 10.6          & 0.235 & 0.0882      & --              & --            & -- \\
    J1846--0513      & recycled & 23.4  & 1.02         & 0.613         & 0.209 & 0.307       & 2.63            & $<1.35$       & $>1.28$ \\
    J1856--0039      & recycled & 23.4  & 0.653        & 0.0985        & 0.106 & 0.104       & 2.49            & $1.30$        & $1.19$ \\ 
    J1901+0658       & recycled & 75.7  & 0.217        & 14.5          & 0.366 & 0.175       & 2.79            & $[1.4]$       & $[1.4]$ \\
    J1906+0746       & young$^*$& 144   & 20300        & 0.166         & 0.0853& 0.112       & 2.61            & 1.29          & 1.32 \\
    J1913+1102       & recycled & 27.3  & 0.161        & 0.206         & 0.0895& 0.137       & 2.89            & 1.62          & 1.27 \\
    B1913+16         & recycled & 59.0  & 8.62         & 0.323         & 0.617 & 0.132       & 2.83            & 1.44          & 1.39 \\
    J1930--1852      & recycled & 186   & 18.0         & 45.1          & 0.399 & 0.348       & 2.59            & $< 1.32$      & $> 1.3$ \\
    J1946+2052       & recycled & 17.0  & 1.11         & 0.0785        & 0.0638& 0.268       & 2.50            & $<1.31$       & $>1.18$ \\
    J2150+3427       & young$^*$& 654   & 3.60         & 10.6          & 0.602 & 0.159       & 2.6             & $[1.3]$       & $[1.3]$ \\
    \hline
    J0514--4002A     & GC$^*$   & 4.99  & 0.0007       & 18.8          & 0.888 & 0.145       & 2.47            & 1.25          & 1.22 \\
    J0514-4002E      & GC$^*$   & 5.60  &  --          & 7.45          & 0.708 & 0.417       & 3.89            & --            & -- \\
    J1748--2446ao    & GC$^*$   & 2.27  & 0.087        & 57.6          & 0.325 & 0.0784      & 3.17            & --            & $> 0.8$ \\
    J1807--2500B     & GC$^*$   & 4.20  & 0.0823       & 9.96          & 0.747 & 0.2620      & 2.57            & 1.37          & 1.21 \\
    J1953+1847D      & GC$^*$   & 101   & 0.080        & 10.9          & 0.629 & 0.2620      & 2.63            & $\sim 1.2$    & $\sim 1.4$ \\
    B2127+11C        & GC$^*$   & 30.5  & 4.99         & 0.335         & 0.681 & 0.153       & 2.71            & 1.36          & 1.35 \\
    J2140--2311B     & GC$^*$   & 13.0  & $-$0.6       & 6.22          & 0.879 & 0.207       & 2.53            & $\leq$1.43    & $\geq 1.10$ \\
\end{tabular}
}
}
\end{table*}

\clearpage
\noindent
{\bf {\Large References}}

\end{document}